\documentclass[fleqn,usenatbib]{mnras}
\usepackage{graphicx} 
\usepackage{txfonts} 
\usepackage{natbib}               
\title[]{The {\it Hubble Space Telescope} UV Legacy Survey of Galactic Globular Clusters. XXI. Binaries among multiple stellar populations.}  
\author[A.\,P.\, Milone et al.] 
       {A.\,P.\,Milone$^{1,2}$, E. Vesperini$^{3}$, A.\,F.\,Marino$^{1,2,4}$, J. Hong$^{3,5}$, R.\,van der Marel$^{6,7}$, \newauthor  J.\,Anderson$^{6}$, 
        A.\,Renzini$^{2}$, G.\,Cordoni$^{1}$, L.\,R.\,Bedin$^{2}$, A.\,Bellini$^{6}$, T.\,M.\,Brown$^{6}$, \newauthor 
        F.\,D'Antona$^{8}$,   E.\,P.\,Lagioia$^{1}$,  M.\,Libralato$^{6}$, D. Nardiello$^{1,2}$, G.\,Piotto$^{1,2}$, %
        M.\,Tailo$^{1}$, \newauthor A. Cool$^{9}$, M.\,Salaris$^{10}$, 
           A.\,Sarajedini$^{11}$ \\ 
$^{1}$Dipartimento di Fisica e Astronomia ``Galileo Galilei'', Univ. di Padova, Vicolo dell'Osservatorio 3, Padova, IT-35122\\
$^{2}$Istituto Nazionale di Astrofisica - Osservatorio Astronomico di Padova, Vicolo dell'Osservatorio 5, Padova, IT-35122\\
$^{3}$Department of Astronomy, Indiana University, Bloomington, IN 47405, USA\\
$^{4}$ Centro di Ateneo di Studi e Attivita' Spaziali ``Giuseppe Colombo'' - CISAS, Via Venezia 15, Padova, IT-35131 \\
$^{5}$ Department of Astronomy, Yonsei University 50 Yonsei-Ro, Seodaemun-Gu, Seoul, South Korea \\
$^{6}$Space Telescope Science Institute, 3800 San Martin Drive, Baltimore,  MD 21218, USA\\
$^{7}$Center for Astrophysical Sciences, Department of Physics \& Astronomy, Johns Hopkins University, Baltimore, MD 21218, USA \\
$^{8}$Istituto Nazionale di Astrofisica - Osservatorio Astronomico di Roma, Via Frascati 33, I-00040 Monteporzio Catone, Roma, Italy\\
$^{9}$ Department of Physics and Astronomy, San Francisco State University, 1600 Holloway Avenue, San Francisco, CA 94132, USA \\
$^{10}$ Astrophysics Research Institute, Liverpool John Moores University, Liverpool Science Park, IC2 Building,
146 Brownlow Hill, Liverpool L3 5RF, UK \\
$^{11}$Department of Astronomy, University of Florida, 211 Bryant Space Science Center, Gainesville, FL 32611, USA \\}

\begin{document} 
\date{Accepted 2019 December 27. Received 2019 December 26; in original form 2019 October 8}
\maketitle 
\label{firstpage}

\begin{abstract}
A number of scenarios for the formation of multiple populations in globular clusters (GCs) predict that second generation (2G) stars form in a compact and dense subsystem embedded in a more extended first-generation (1G) system. 
 If these scenarios are accurate, a consequence of the denser 2G formation environment is that 2G binaries should be more significantly affected by stellar interactions and disrupted at a larger rate than 1G binaries.
 The fractions and properties of binary stars can thus provide a dynamical fingerprint of the formation epoch of multiple-population GCs and their subsequent dynamical evolution.
 We investigate the connection between binaries and multiple populations in five GCs, NGC\,288, NGC\,6121 (M\,4), NGC\,6352, NGC\,6362, and NGC\,6838 (M\,71).
 To do this, we introduce a new method based on the comparison of {\it Hubble Space Telescope} observations of binaries in the F275W, F336W, F438W, F606W and F814W filters with a large number of simulated binaries.
 In the inner regions probed by our data we do not find large differences between the local 1G and the 2G binary incidences in four of the  studied clusters, the only exception being M\,4 where the 1G binary incidence is about three times larger than the 2G incidence. The results found are in general agreement with the results of simulations  predicting significant differences in the global 1G and 2G incidences and in the local values in the clusters' outer regions but similar incidences in the inner regions.  
 The significant difference found in M\,4 is consistent with simulations with a larger fraction of wider binaries.
 Our analysis also provides the first evidence of mixed (1G-2G) binaries, a population predicted by numerical simulations to form in a cluster's inner regions as a result of stellar encounters during which one component of a binary is replaced by a star of a different population.
\end{abstract}

 
\section{Introduction}\label{sec:intro}
Binary stars play a key role in many aspects of globular clusters' (GCs) dynamics and their evolution and survival is, in turn, significantly affected by stellar interactions in the clusters' dense environment \citep[see e.g.\,][]{heggie2003a}.

A variety of scenarios predict that 2G stars formed in a  high-density environment in the cluster center \citep[e.g.][]{dercole2008a, calura2019a}. 
 Since the rates of binary disruption and evolution of the parameters of surviving binaries strongly depend on the stellar density, the incidence of binaries in first- and second-generation (hereafter 1G and 2G) stars of GCs can provide information and constraints on their formation environment and their long-term evolution \citep{vesperini2011a, hong2015a, hong2016a, hong2019a}.

  Indeed a number of numerical studies have shown that the {\it global} present-day incidence of binaries in the 2G population is expected to be lower than that of  1G stars \citep{vesperini2011a, hong2015a, hong2016a}. This is a consequence of the larger effect of dynamical processes that determine the evolution and disruption of binary stars for the more concentrated 2G population. In the interpretation of observations covering a specific range of radial distances from a cluster's center, it is necessary to consider that {\it local} values of the 1G and 2G binary incidences (i.e. values of the binary incidence measured at a given distance from a cluster's center) are determined by a combination of dynamical effects on binary evolution and disruption and the extent of spatial mixing reached by a cluster at any given time during its dynamical evolution \citep{hong2019a}.



The first attempts to infer the incidence of binaries in multiple populations 
were based on spectroscopy. 
On the basis of a study of 21 radial-velocity (RV) binaries in ten GCs, 
\citep{lucatello2015a} concluded that the fraction of binaries among 1G is 4.1$\pm$1.7 times higher than the fraction of binaries in the 2G \citep[see also][]{dorazi2010a}.  More recently, 
\citet{dalessandro2018a} found that only one out of twelve RV binaries 
in the GC NGC\,6362 belong to the 2G population.  This corresponds to 
a fraction of binaries in the 1G and 2G populations equal to, respectively,  4.7$\pm$1.4\% and 0.7$\pm$0.7\%.
These studies probed mainly  clusters' regions around the half-light radius and the differences found between the 1G and 2G binary incidences revealed a larger 1G binary incidence in general agreement with the theoretical expectations.

In the analysis presented here, we exploit multi-band {\it Hubble Space Telescope} ({\it HST}) photometry collected as part of the UV Legacy Survey of Galactic GCs \citep{piotto2015a} to study binaries among multiple populations in five GCs, namely NGC\,288, NGC\,6121, NGC\,6352, NGC\,6362 and NGC\,6838.
These GCs are all relatively simple objects in the context of multiple populations and share three properties that make them ideal targets to investigate the incidence of binaries among 1G and 2G stars.
\begin{itemize}
\item Their 1G and 2G stars exhibit moderate variations in  their chemical composition, yet even so the two populations are still distinct \citep[e.g.][]{marino2008a, marino2011a, carretta2009a}. This is in contrast with massive GCs, where 1G and 2G stars host  sub-populations with large differences in helium and light-element abundance \citep[e.g.][]{milone2017a, milone2018a, marino2019a}.
\item The two distinct groups of 1G and 2G stars are well separated along the main sequence (MS), sub-giant branch (SGB), and red-giant branch (RGB) either in the chromosome map (ChM) or in appropriate color-color diagrams.
\item 1G and 2G stars are distinguishable in the ChMs of MS stars that are at least two magnitudes fainter than the MS turn off in the F814W band. 
\end{itemize}
  
This paper is organized as follows.  In Section~\ref{sec:data} we 
describe the data and the data reduction. The multiple populations of each cluster are discussed in Section~\ref{sec:mpops}, where we identify the 
two groups of single 1G and 2G stars along the CMD. Section~\ref{sec:bin} 
is dedicated  to the presentation of the results and a discussion of the connection between binaries and multiple populations. Finally, discussion and conclusions are provided in 
Section~\ref{sec:discussion} and \ref{sec:conclusions}, respectively.

\section{Data and data analysis} \label{sec:data} 
%
%
The dataset used in this paper consists of images collected through 
the Wide Field Channel of the Advanced Camera for Survey (WFC/ACS) 
and the Ultraviolet and Visual Channel of the Wide Field Camera 3 
%
%
(UVIS/WFC3) on board {\it HST}. The main properties of the images 
are summarized in \citet{piotto2015a} and \citet{milone2018a}.

%
%
To derive the photometry and the astrometry of all the stars we used 
the FORTRAN software package KS2 developed by Jay Anderson, 
\citep[see][for details]{sabbi2016a,bellini2017a,nardiello2018b} KS2 is the evolution of $kitchen\_sync$, originally 
developed by \citet{anderson2008a} to reduce two-filter WFC/ACS 
globular cluster data.

KS2 uses different methods to measure stars with different brightnesses.
Fluxes and positions of the bright stars were fit for position and flux 
in each individual exposure independently using the best point-spread function (PSF) model for the 
star's location on the detector.  The various measurements of each star 
were then averaged to derive the best estimates of stellar magnitude 
and position.

Faint stars often do not have enough flux to measure their magnitudes 
and positions in individual exposures.  Hence, the KS2 routine determines 
for each star an average position from all the exposures, then it fits 
each exposure's pixels with the PSF, solving only for the flux.   

Stellar positions have been corrected for geometrical distortion by using 
the solutions by \citet{bellini2009a} and \citet{bellini2011a}.  The photometry has been converted from the instrumental system into the Vega system as in \citet{bedin2005a} using the updated zero points of the WFC/ACS and UVIS/WFC3 filters available at the STScI web pages.

We used the diagnostics of the photometric and astrometric qualities provided by KS2
 to select a sample of relatively isolated stars that are well fitted by the PSF. 
  Specifically, we exploited position and magnitude rms, the fraction of flux in the aperture due to neighbours and the quality of the PSF fit.
 We plotted each parameter as a function of the stellar magnitude and verified that most stars follow a clear trend in close analogy with what is done in previous papers from our group \citep[e.g.\,][]{milone2009a, bedin2009a}.
  Outliers include variable stars and stars with poor astrometry and photometry and are excluded from our investigation. 

The fluxes of stars in the field of view of NGC\,6121, NGC\,6352, NGC\,6362 and NGC\,6838 are significantly affected 
by spatial variation of the interstellar extinction.  To minimize the 
artificial broadening of the photometric sequences in the CMDs due 
to spatial variations of the photometric zero points, the photometry has 
been corrected for differential reddening using the procedure by 
\citet{milone2012a}.

 \subsection{Artificial stars}
To derive the fraction of 1G and 2G stars among the binaries, we 
compared the observed photometric diagrams of each GC with simulations, 
which are constructed from artificial-star (AS) photometry.
%
%
AS tests have been run by following the method by \citet{anderson2008a}. 
In a nutshell, we generated a catalog of 300,000 stars with instrumental 
F814W magnitude from the saturation limit of the images to  the instrumental magnitude $-4.0$, which is below the detection threshold of our data.  Instrumental 
magnitudes are defined as $-2.5 \cdot \log_{10}{\rm (flux)}$, where 
the flux is given in photo-electrons.
 
The F275W, F336W, F438W, and F606W magnitudes were calculated from 
the colors of the fiducials lines of 1G and 2G stars, which are 
derived from the observed CMDs.  We associated to each AS a position 
in such a way that the radial distribution of ASs resembles the 
radial distribution of stars brighter than $m_{\rm F814W}=21.0$.
ASs were reduced using the same method adopted for real stars and we 
included in our analysis only those ASs that pass the criteria of 
selection used for real stars.  The ASs were inserted, found, and 
detected one at a time, so that they would never interfere with each
other.

\section{Multiple stellar populations}
\label{sec:mpops}
\begin{centering} 
\begin{figure*} 
  \includegraphics[width=12.0cm,trim={0cm 8.cm 0.2cm 1.cm},clip]{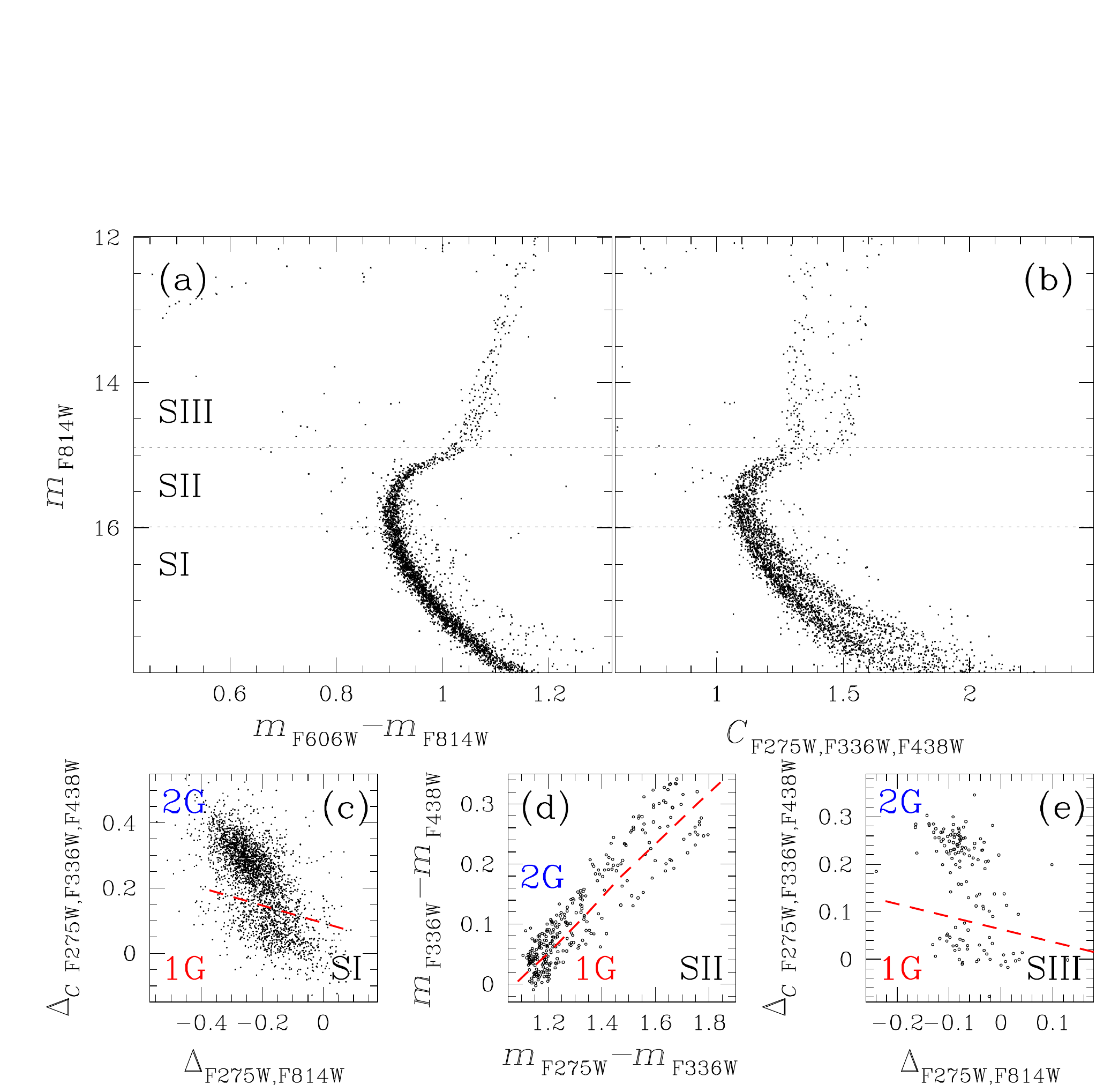}
  \includegraphics[width=12.0cm,trim={0cm 14.2cm 0.2cm 5.cm},clip]{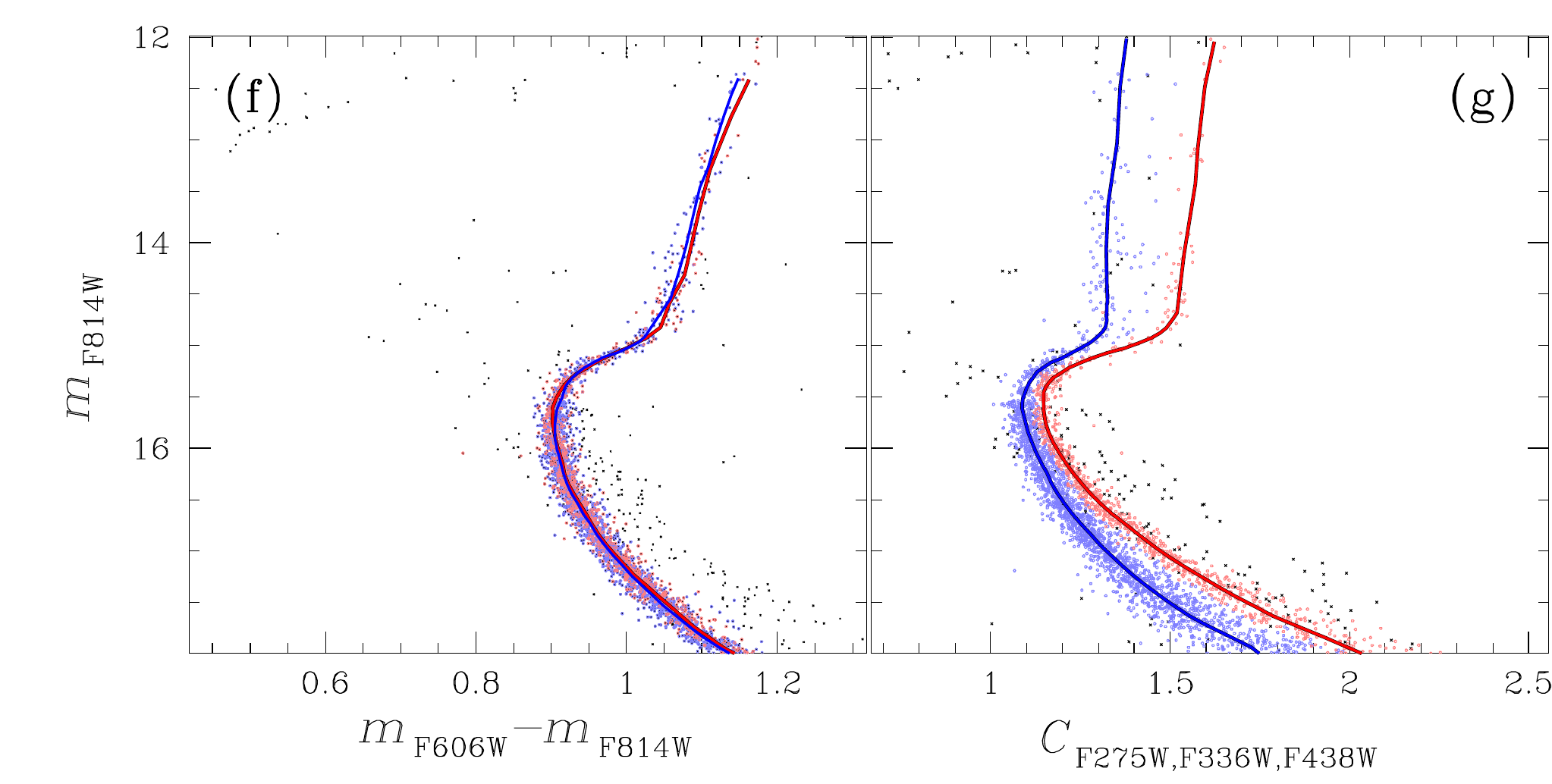}
  \caption{This figure summarizes the method that we used to select 
           1G and 2G stars along the CMD of M\,4. Panels (a) and (b) 
           show the $m_{\rm F814W}$ vs.\,$m_{\rm F606W}-m_{\rm F814W}$ 
           CMD and the $m_{\rm F814W}$ vs.\,$C_{\rm F275W,F336W,F438W}$ 
           pseudo-CMD of M\,4. Panels (c), (d), and (e) show either the 
           chromosome map or the $m_{\rm F336W}-m_{\rm F438W}$ 
           vs.\,$m_{\rm F275W}-m_{\rm F336W}$ two-color diagram of stars 
           in the regions SI, SII, and SIII of the CMD indicated in 
           panels (a) and (b). The red-dashed lines are used to separate 
           1G from 2G stars. Panels (f) and (g) reproduce the diagrams 
           plotted in panels (a) and (b). 1G and 2G stars are colored 
           red and blue, respectively, and the corresponding fiducial 
           lines are superposed on the diagrams.  }
 \label{fig:sel} 
\end{figure*} 
\end{centering}

As a first step to study the binaries among multiple populations
we identified 1G and 2G stars along the MS, the SGB, and the RGB.  To 
do this, we adopted the procedure illustrated in Figure~\ref{fig:sel} for M\,4, 
which is based on photometric diagrams that maximize the separation 
between stellar populations with different chemical compositions. 

We used different diagrams to identify 1G and 2G stars at different brightness
levels.  Specifically, we defined the three intervals of F814W magnitude, 
SI, SII, and SIII, which are indicated by the dotted lines in the 
$m_{\rm F814W}$ vs.\,$m_{\rm F606W}-m_{\rm F814W}$ and the 
$m_{\rm F814W}$ vs.\,$C_{\rm F275W,F336W,F438W}$ diagrams plotted in 
panels (a) and (b) of Figure~\ref{fig:sel}.  Due to the large observational 
errors, we are not able to clearly distinguish 1G and 2G stars below 
$m_{\rm F814W}=18.5$.

%
%
Panels (c) and (e)  of Figure~\ref{fig:sel} show that the distribution 
of SI and SIII stars in the $\Delta_{C{\rm F275W,F336W,F438W}}$ 
vs.\,$\Delta_{\rm F275W,F814W}$ pseudo two-color diagram, otherwise known
as a chromosome map \citep[][]{milone2015a, milone2017a} is bimodal. 
Similarly, the SII stars are distributed along two sequences in the 
$m_{\rm F336W}-m_{\rm F438W}$ vs.\,$m_{\rm F275W}-m_{\rm F336W}$ 
two-color diagram, in close analogy with what we have observed in other 
GCs \citep[][]{milone2012c, milone2013a, tailo2019a}.  The red lines, which are drawn 
by hand with the aim of separating the two main stellar sequences 
within each diagram, are used to define the populations of 1G and 2G 
stars.

%
%
1G and 2G stars, selected in panels (c), (d), and (e) are colored red 
and blue, respectively, in the diagrams plotted in panels (f) and (g). 
The red and the blue lines superposed on each diagram are the fiducials 
of 1G and 2G stars.  To derive these lines we used a method that is 
based on the naive estimator by \citet{silverman1986a}. In a nutshell, we first 
defined a series of magnitude intervals of width $\nu$, from 
$m_{\rm F814W}=12.0$ to $18.5$. We used $\nu$=0.2, 0.1, and 0.4 for stars 
in the SI, SII, and SIII regions of the CMD. These intervals are defined 
over a grid of points separated by steps of fixed magnitude ($s=\nu/3$). 
For each interval we calculated the median color and magnitude and 
smoothed these median points by boxcar averaging, where each point 
is replaced by the average of the three adjacent points.

\begin{centering} 
\begin{figure*} 
  \includegraphics[width=8.75cm,trim={0cm 5.cm 0.2cm 5.cm},clip]{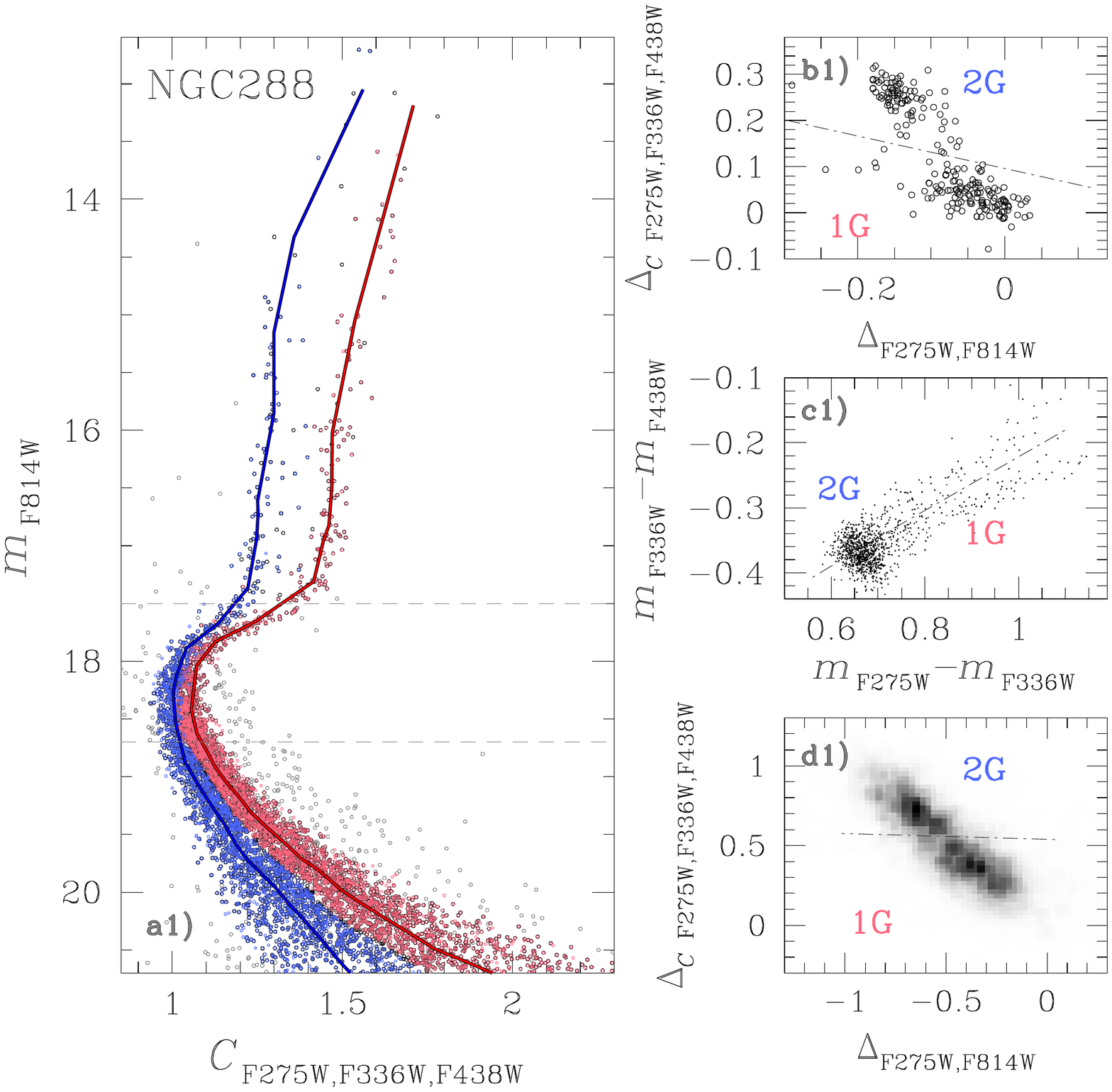}
  \includegraphics[width=8.75cm,trim={0cm 5.cm 0.2cm 5.cm},clip]{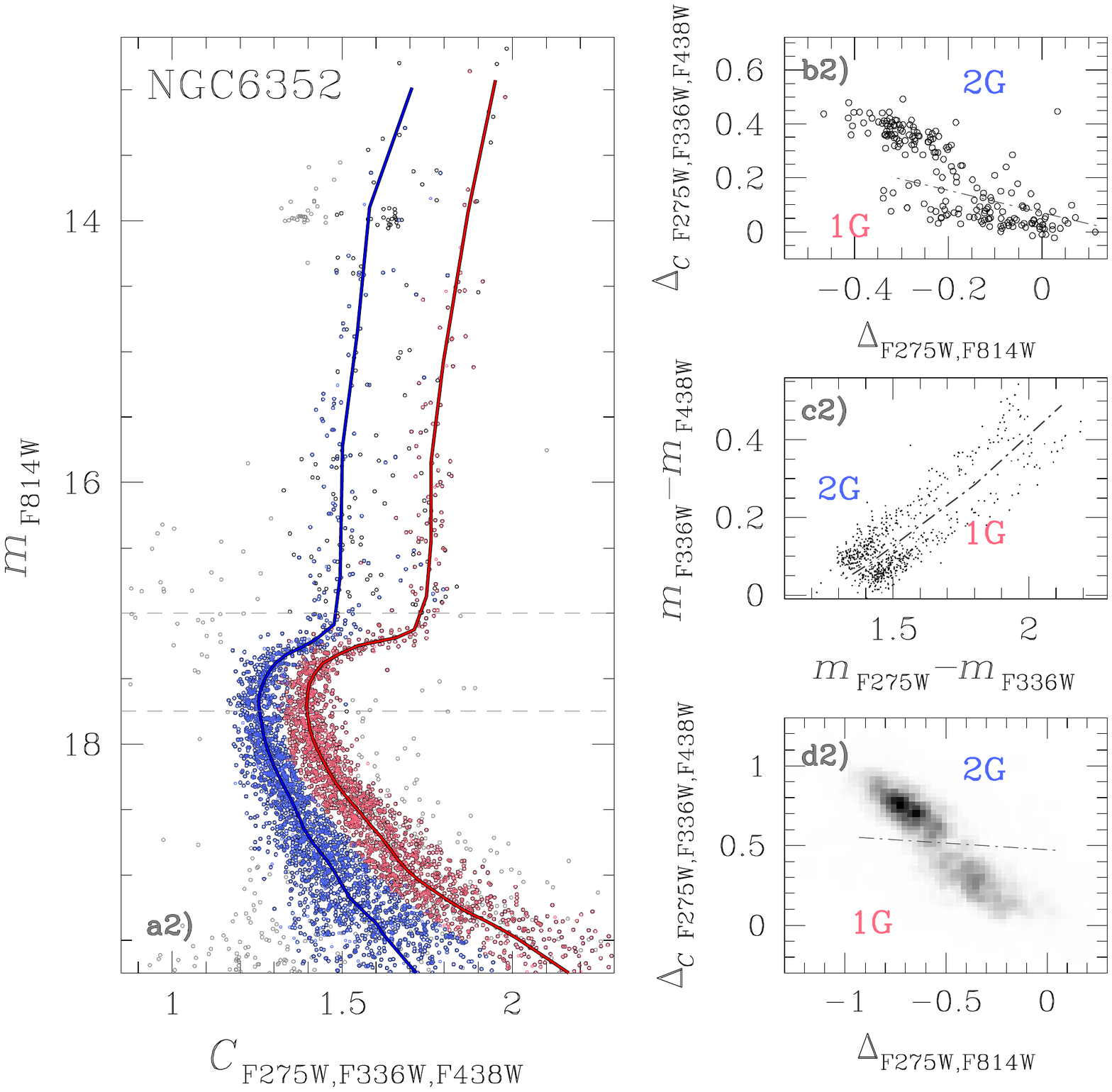}
  \includegraphics[width=8.75cm,trim={0cm 5.cm 0.2cm 4.9cm},clip]{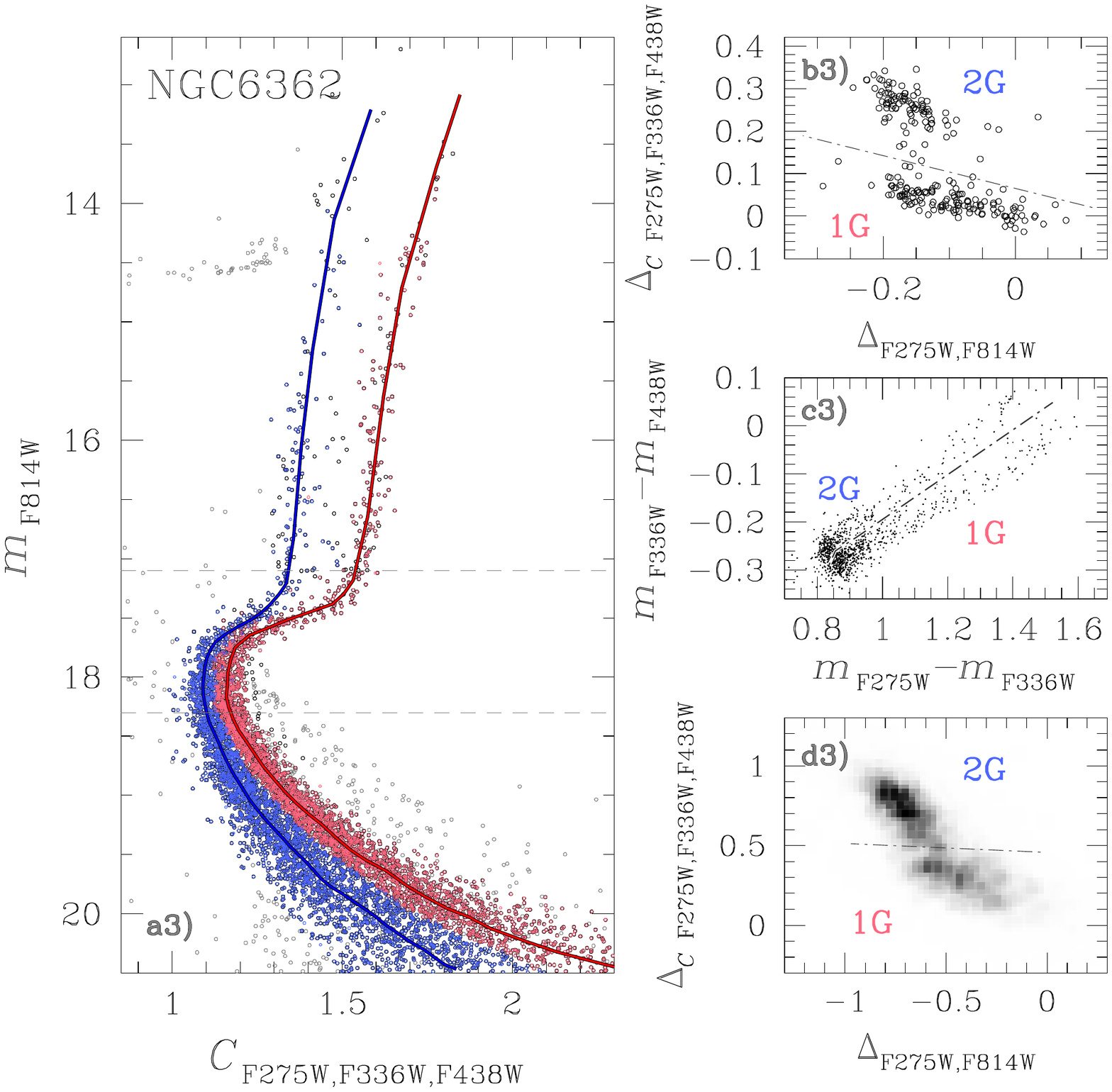}
  \includegraphics[width=8.75cm,trim={0cm 5.cm 0.2cm 4.9cm},clip]{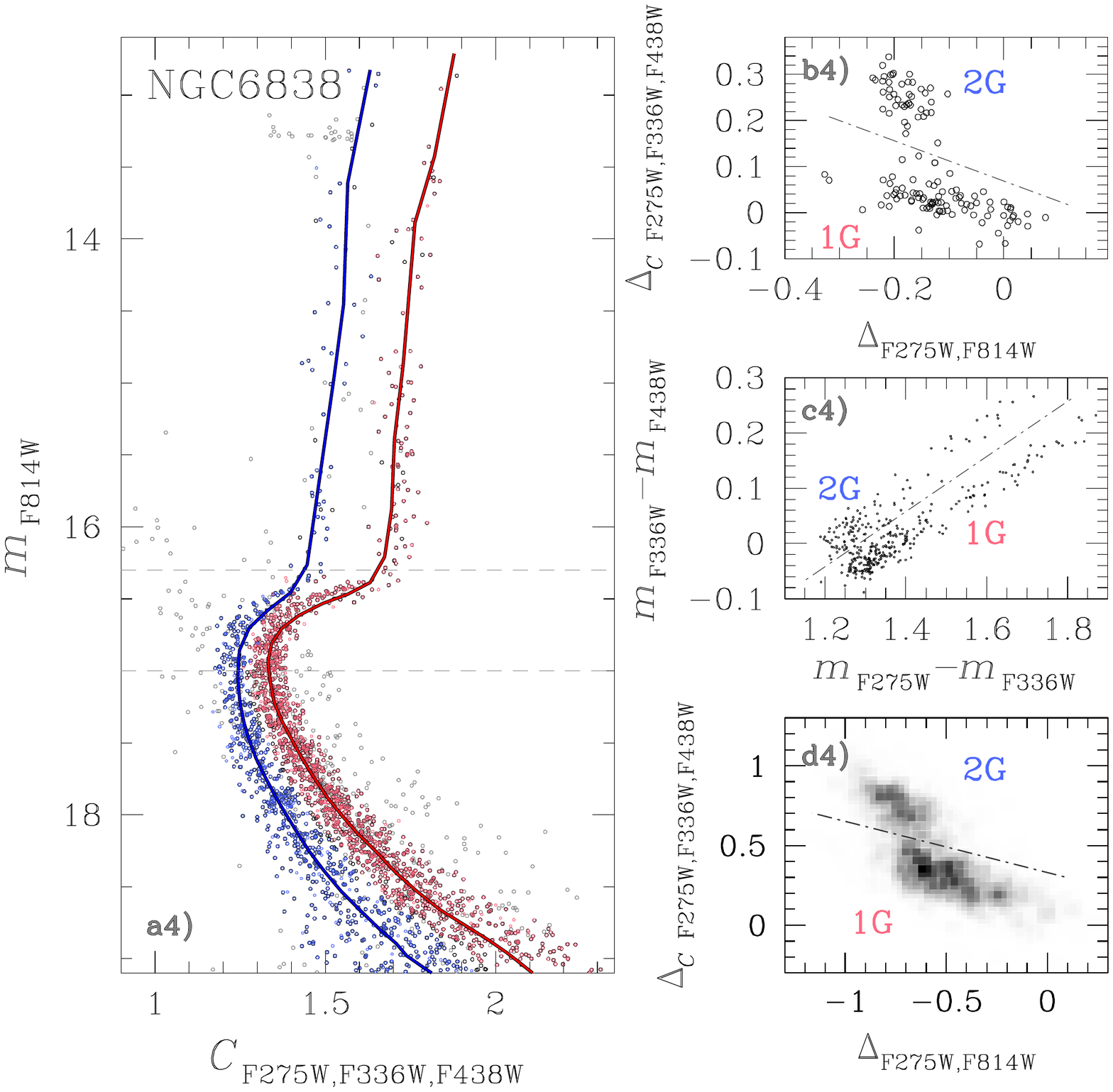}
  \caption{The red and blue colors mark the selected 1G and 2G stars, respectively, in the $m_{\rm F814W}$ vs.\,$C_{\rm F275W,F336W,F438W}$ diagrams (panels a1--a4) for NGC\,288, NGC\,6352, NGC\,6362 and NGC\,6838. The horizontal dashed lines separate the SI, SII, and SIII regions of the diagram that mostly populated by RGB, SGB, and MS stars, while the red and blue lines superimposed on the diagram are the fiducial lines of 1G and 2G stars.
     Panels  b1--b4, c1--c4, d1--d4 show the ChM of RGB stars, the $m_{\rm F336W}-m_{\rm F438W}$  vs.\,$m_{\rm F275W}-m_{\rm F336W}$ two-color diagram of SGB stars, and the Hess diagram of the MS ChM. The dashed-dot lines separate 1G and 2G stars.}
 \label{fig:sel2} 
\end{figure*} 
\end{centering}

We followed the procedure described above for M\,4 to identify 1G and 2G stars along the RGB, SGB and MS of the other studied GCs.
Results are summarized in Figure~\ref{fig:sel2} where we use red and blue colors to represent 1G and 2G stars, respectively, in the $m_{\rm F814W}$ vs.\,$C_{\rm F275W,F336W,F438W}$ diagram (panels a1--a4). We also show the ChMs of RGB and MS stars and the $m_{\rm F336W}-m_{\rm F438W}$  vs.\,$m_{\rm F275W}-m_{\rm F336W}$ two-color diagram of SGB stars  that we used to select 1G and 2G stars, in close analogy with what we did for M\,4.  The ChMs of MS stars are used to obtain the fractions of 1G stars of each clusters that listed in Table~\ref{tab:res} and are derived as in \citet{milone2017a}. 

\section{Binaries and multiple populations}\label{sec:bin}
The binary systems that survive in the dense environment of a GC are the 
extremely tight ones. For this reason, their individual components are not 
resolved in the {\it HST} images and the binary system appears in our
images as a single point source.  The position in the CMD of a binary 
system formed by non-interacting stars is related to the luminosity of 
its two components.  Specifically, the magnitude of the binary system is:
\begin{equation}
  m_{\rm bin}=m_{1}-2.5 \log {\Big{(}1+\frac{F_{2}}{F_{1}}\Big{)}}
\end{equation}
where $F_{1}$ and $F_{2}$ are the fluxes of the two stars and 
$m_{1}=-2.5 \log{F_{1}} + constant $.

\begin{centering} 
\begin{figure*} 
  \includegraphics[width=8.75cm,trim={0cm 5.cm 2.2cm 10.cm},clip]{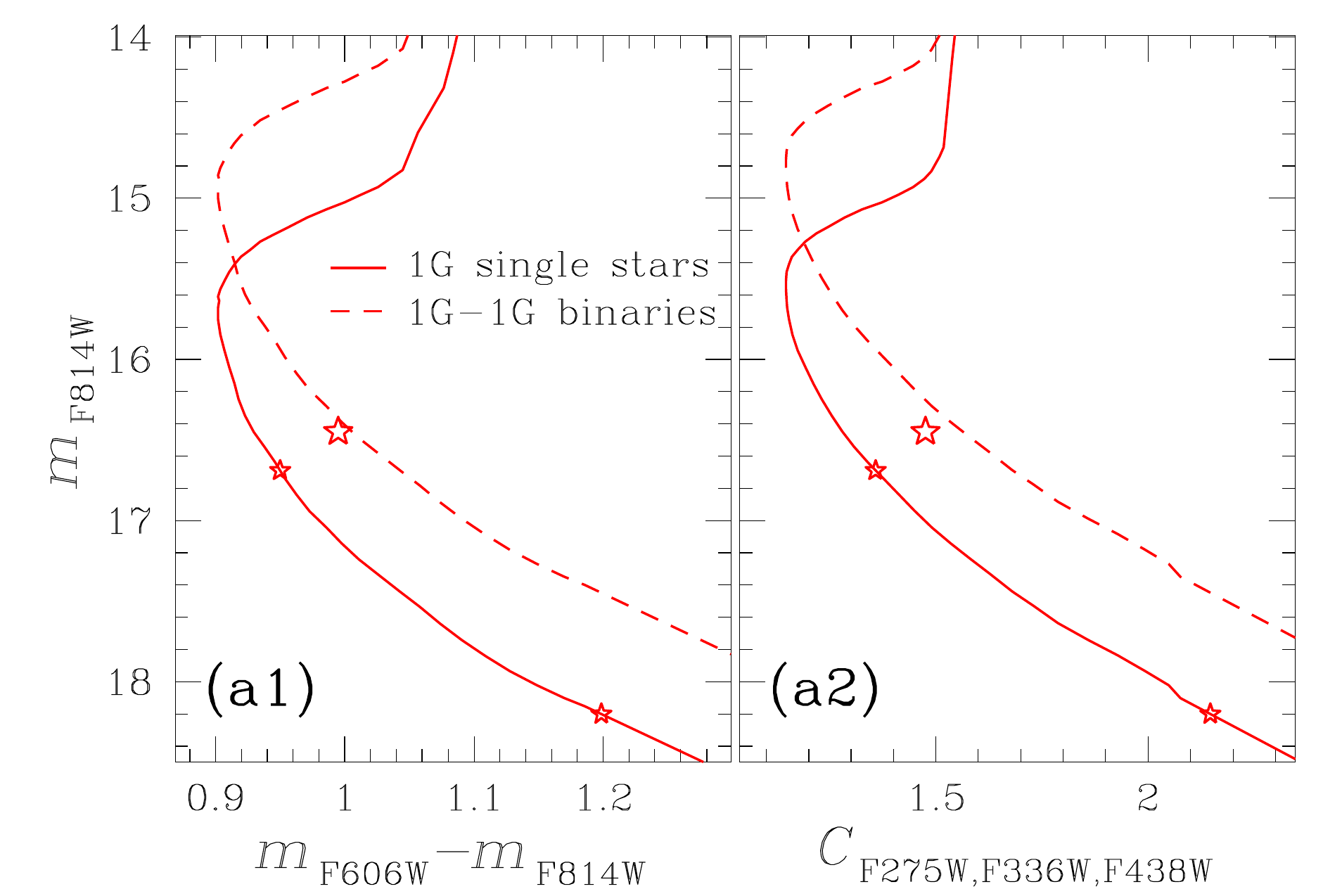}
  \includegraphics[width=8.75cm,trim={0cm 5.cm 2.2cm 10.cm},clip]{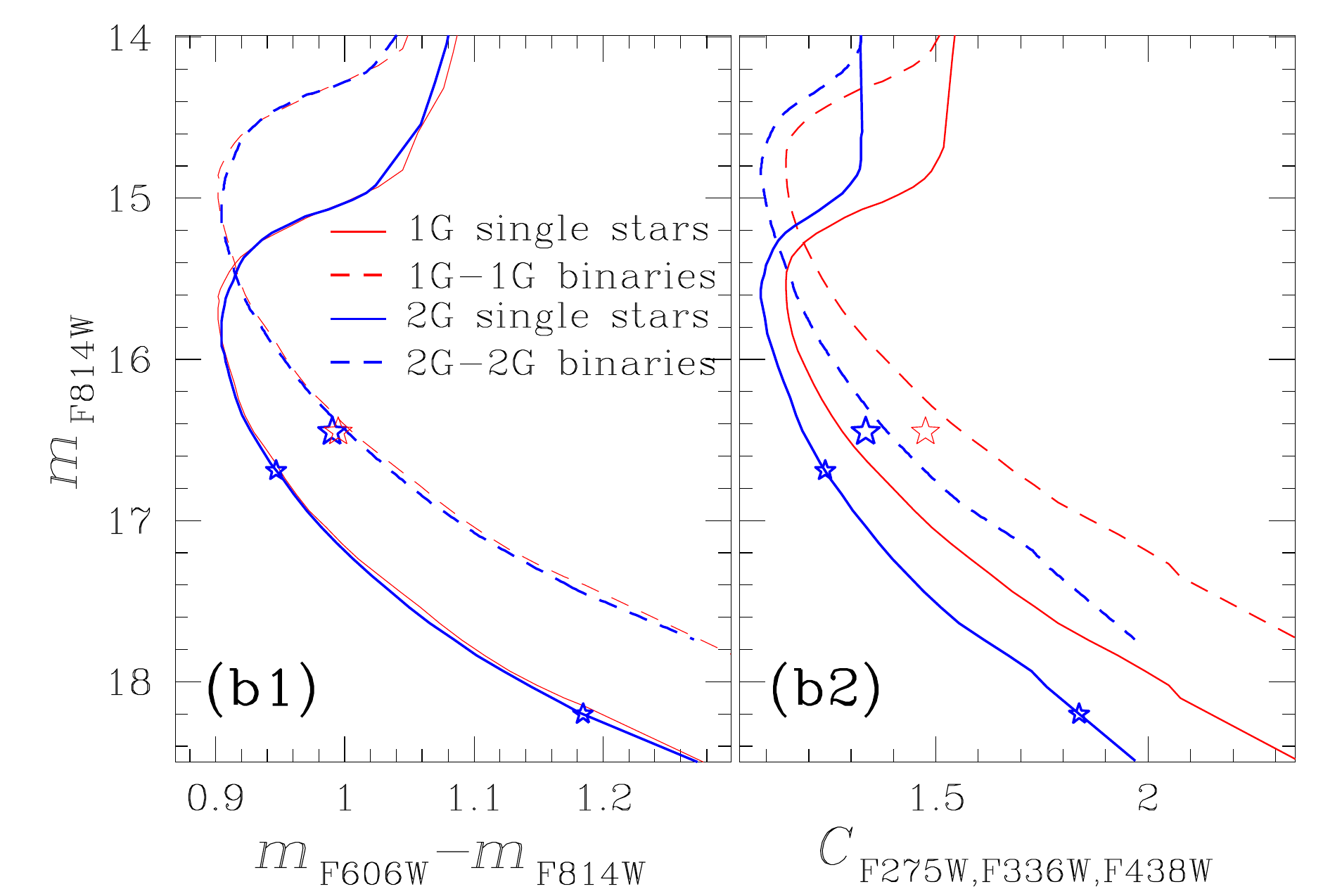}
  \includegraphics[width=8.75cm,trim={0cm 5.cm 2.2cm 10.cm},clip]{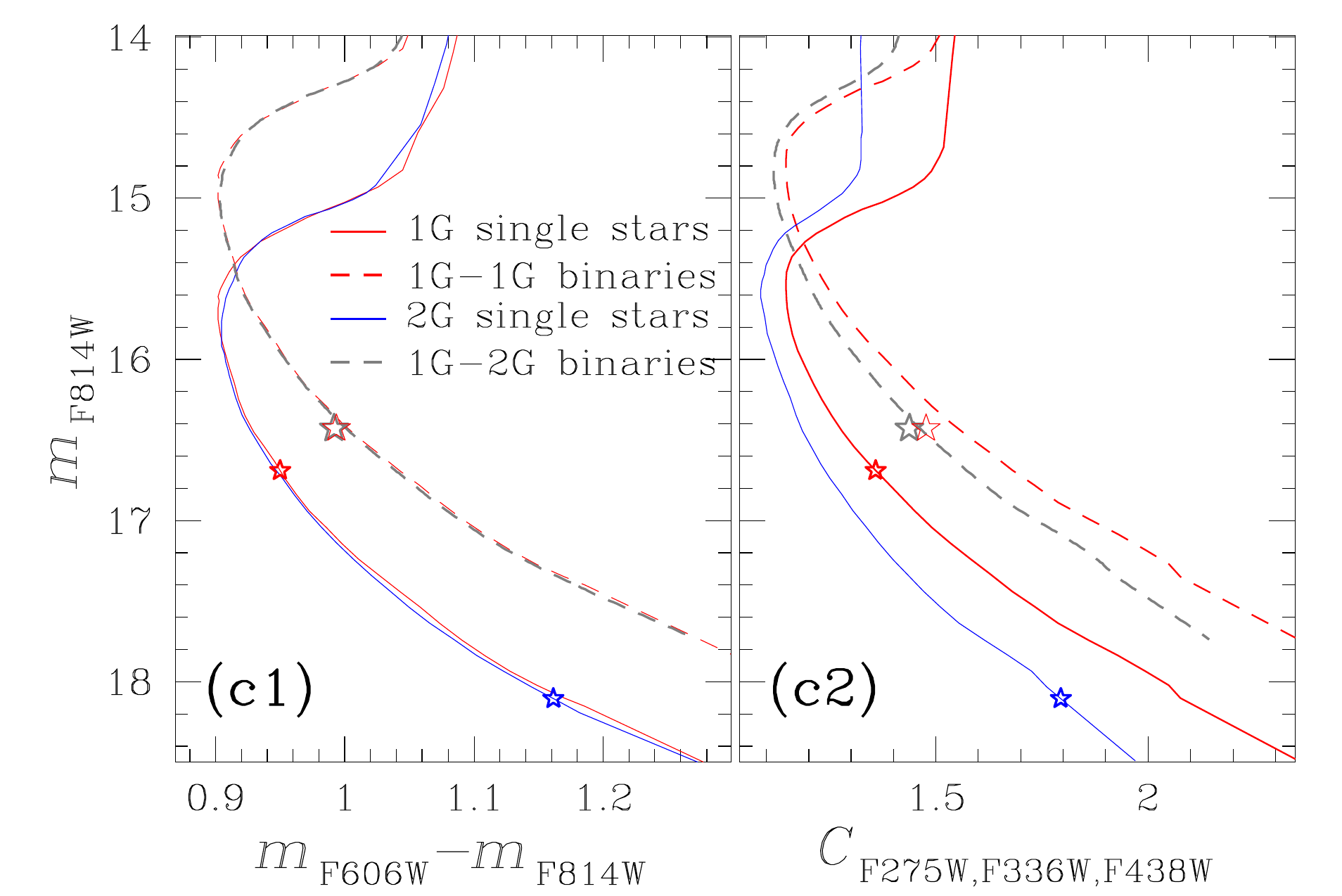}
  \includegraphics[width=8.75cm,trim={0cm 5.cm 2.2cm 10.cm},clip]{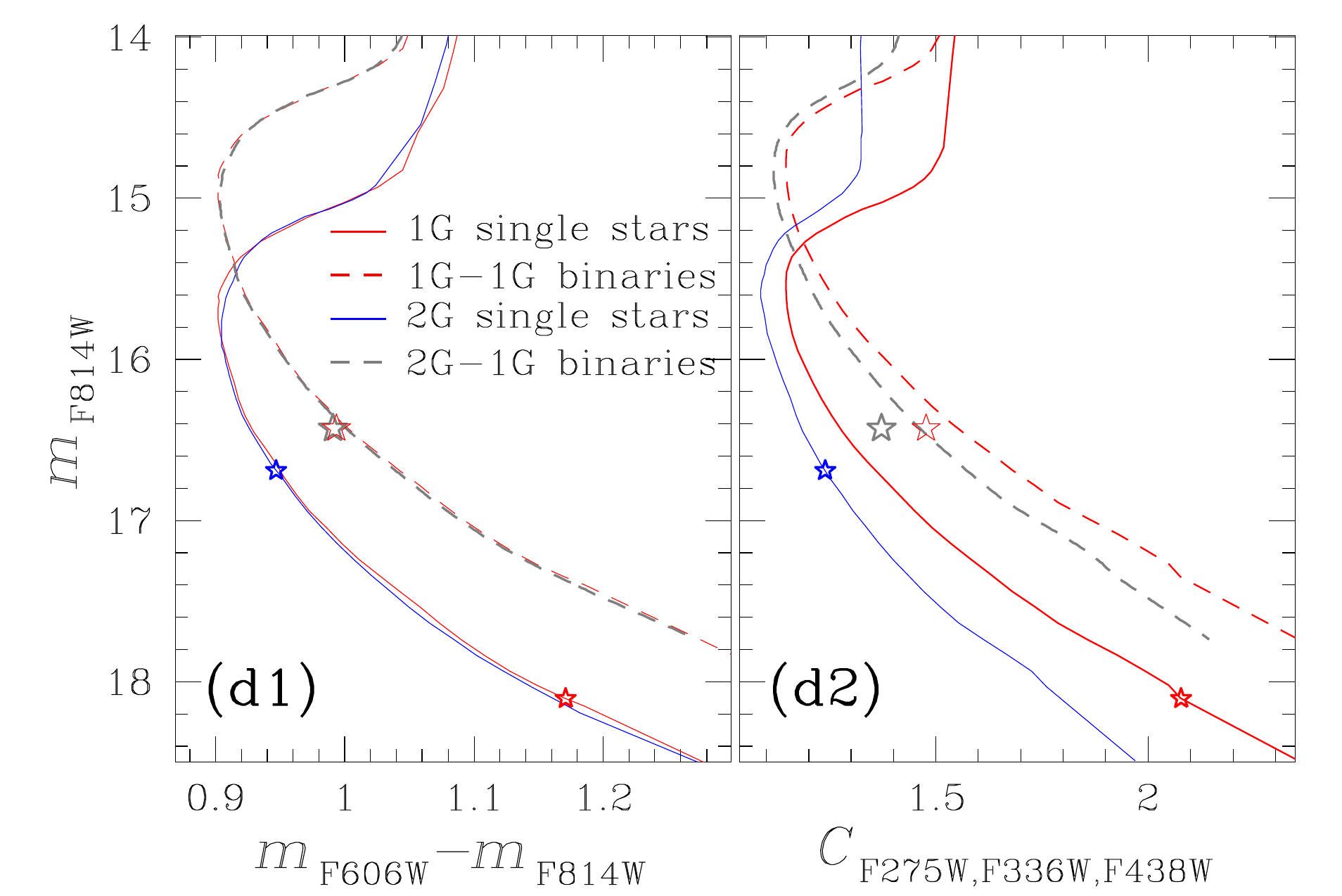}
  \caption{The red and blue continuous lines are the fiducials of 
           single 1G and 2G stars, respectively in the $m_{\rm F814W}$ 
           vs.\,$m_{\rm F606W}-m_{\rm F814W}$ CMDs (panels a1, b1, c1, 
           and d1) and $m_{\rm F814W}$ vs.\,$C_{\rm F275W,F336W,F438W}$ 
           pseudo-CMDs (panels a2, b2, c2, and d2).  The fiducial lines 
           of binary pairs made of two stars with the same F814W luminosity 
           are represented with dashed lines. Specifically red, blue, 
           gray and blue dashed lines represent binaries formed by 
           1G-1G, 2G-2G, 1G-2G, 2G-1G stars.  The large starred symbols 
           indicate a binary formed by two MS stars with 
           $m_{\rm F814W}$=$16.7$ and $m_{\rm F814W}$=$18.2$ (small 
           starred symbols).  Specifically, in panels a1-a2 and b1-b2 both components of 
           this binary system are 1G stars (small red starred symbols) and     
           2G stars (small blue starred symbols), respectively,
           while in panels c1-c2 and d1-d2 we combined 1G and 2G stars.
            For comparison purpose, we plot in each 
           panel the binary system represented with the red large 
           starred symbol in panels (a1) and (a2).} 
 \label{fig:teo} 
\end{figure*} 
\end{centering}

In the case of a simple stellar population, the binaries formed by two stars with the same luminosity form a sequence that runs parallel to the cluster fiducial line but is $\sim$0.75 mag brighter. Binaries formed by stars with different luminosities will populate the region of the CMD delimited by the fiducial lines of single stars and the equal-mass binaries. 
%
%
In panels (a1) and (a2) of Figure~\ref{fig:teo}, we plot with continuous 
red lines the fiducials of 1G stars in the $m_{\rm F814W}$ 
vs.\,$m_{\rm F606W}-m_{\rm F814W}$ CMD and in the $m_{\rm F814W}$ 
vs.\,$C_{\rm F275W,F336W,F438W}$ pseudo-CMD, respectively. The fiducials 
of binaries formed by two 1G stars with the same luminosity are represented 
with red dashed lines.  
To illustrate the behaviour of a binary system composed of stars with different luminosities, we represent with a large 
red-starred symbol the binary system formed by two 1G MS stars with 
$m_{\rm F814W}$=$16.7$ and $m_{\rm F814W}$=$18.2$ whose components 
are indicated with small red-starred symbols.  The fiducials and the 
binary stars introduced in panels (a1) and (a2) are reproduced in all 
the panels of Figure~\ref{fig:teo}.

In panels (b1) and (b2) of Figure~\ref{fig:teo} we represent with 
blue continuous and dashed lines the fiducials of single 2G MS stars 
and of 2G-2G equal-luminosity binaries, respectively. 2G-2G binaries have 
similar $m_{\rm F606W}-m_{\rm F814W}$ colors as 1G-1G binaries with the 
same luminosity but substantially different values of 
$C_{\rm F275W,F336W,F438W}$.   

In the bottom panels of Figure~\ref{fig:teo} we considered binaries 
formed by 1G and 2G stars and we used gray colors to represent the 
fiducials of equal-luminosity binaries. In panels (c1) and (c2) the 
brightest component of all the binary systems belong to the 1G while in 
panels (d1) and (d2) the 2G star is brighter than its 1G companion. 
For fixed F814W magnitudes of 1G and 2G stars, the latter case results 
in smaller values of $C_{\rm F275W,F336W,F438W}$. In general, binaries 
formed by 1G and 2G pairs have $C_{\rm F275W,F336W,F438W}$ values 
that are in between those of the 1G-1G and 2G-2G binaries.  

%
\subsection{The sample of binaries}
\begin{centering} 
\begin{figure*} 
  \includegraphics[height=8.5cm,trim={0cm 5.cm 0.2cm 4.45cm},clip]{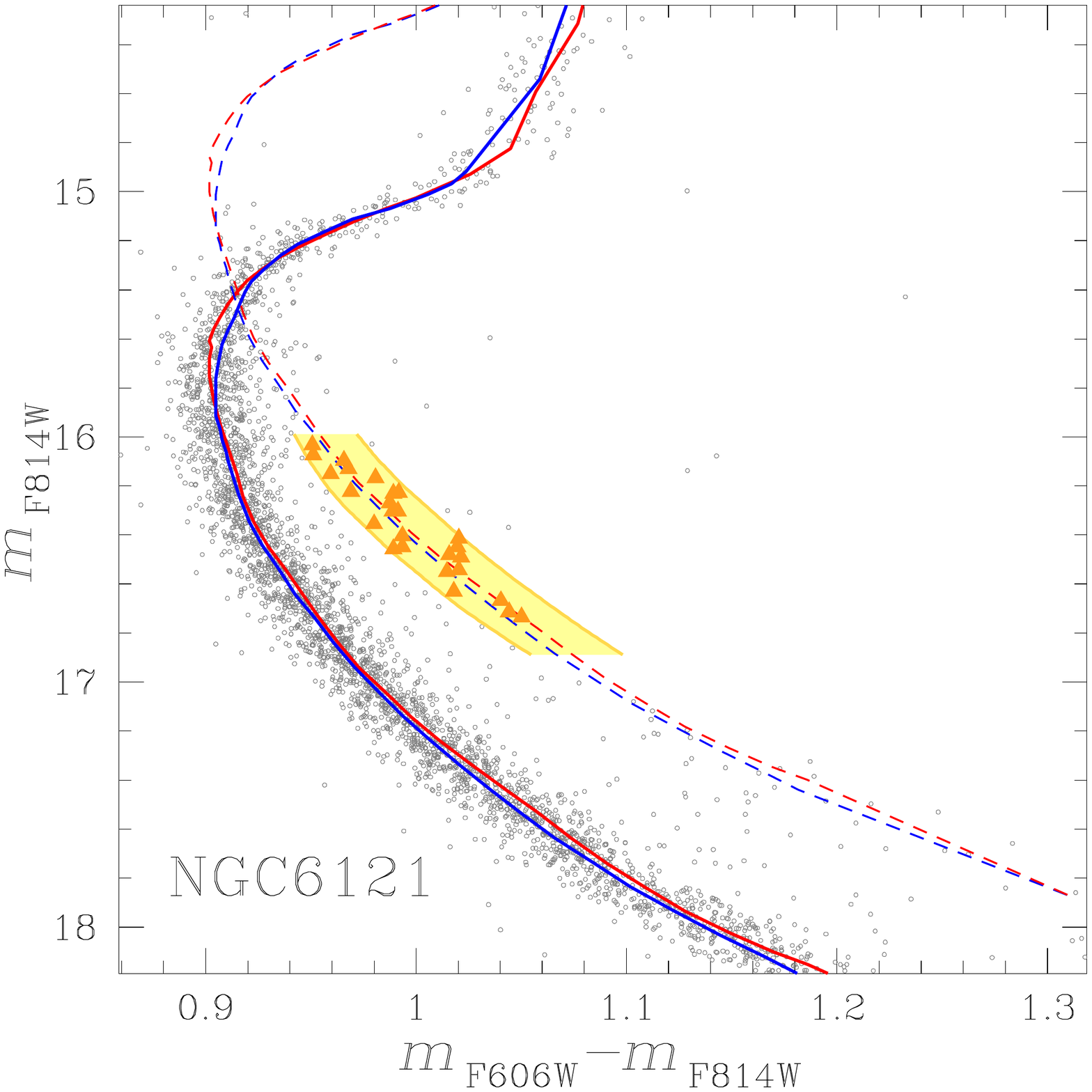}
  \includegraphics[height=8.5cm,trim={0.8cm 5.cm 0.2cm 2.7cm},clip]{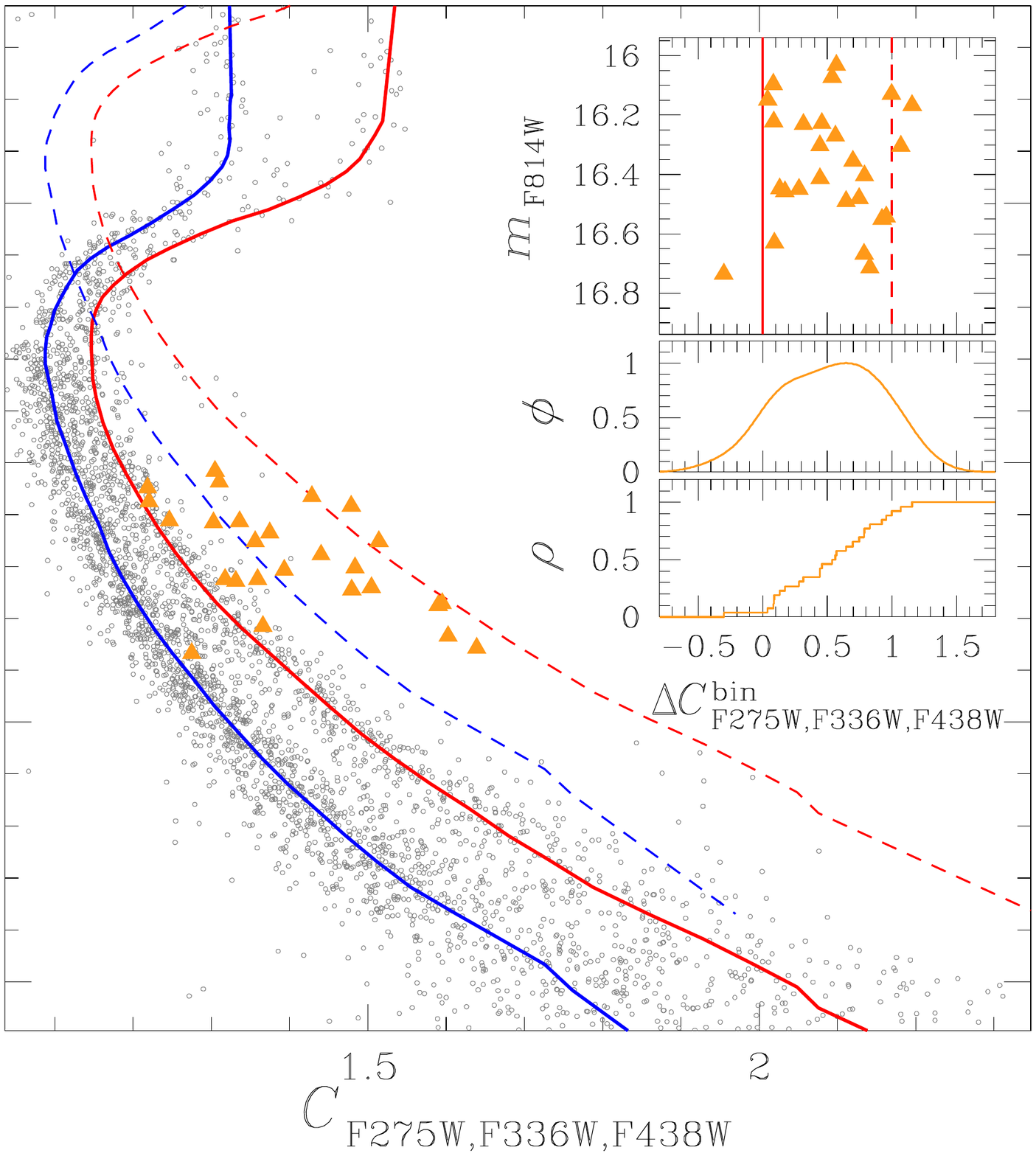}
  \caption{$m_{\rm F814W}$ vs.\,$m_{\rm F606W}-m_{\rm F814W}$ CMD 
           (left panel) and $m_{\rm F814W}$ 
           vs.\,$C_{\rm F275W,F336W,F438W}$ diagram of 
           M\,4 (right panel).  The fiducial lines of 1G and 2G stars 
           are plotted with continuous lines, while the dashed lines 
           represent fiducials for equal-mass binaries formed by pairs 
           of 1G and 2G stars.  Red and blue colors refer to 1G and 2G, 
           respectively.  The binaries that will be investigated in this 
           analysis are marked with orange triangles, and are selected 
           from the left-panel CMD.  The inset shows the verticalized 
           $m_{\rm F814W}$ vs.\,$\Delta C_{\rm F275W,F336W,F438W}$ 
           diagram for the selected binaries (top), the corresponding 
           $\Delta C_{\rm F275W,F336W,F438W}$ kernel distribution (middle), and the cumulative distribution (bottom). 
           See text for details.} 
 \label{fig:selBIN} 
\end{figure*} 
\end{centering}

The binaries of M\,4 analyzed in this paper are located in the shaded yellow  region of the $m_{\rm F814W}$ vs.\,$m_{\rm F606W}-m_{\rm F814W}$ CMD  plotted in the left panel of Figure~\ref{fig:selBIN}, which is delimited by the two yellow segments: the segment with the reddest color is the  fiducial of the equal-mass 1G-1G binaries but shifted to the red by two  times the $m_{\rm F606W}-m_{\rm F814W}$ color error.  The other yellow  segment is the fiducial formed by a binary system that includes one 
2G star with $m_{\rm F814W}=18.5$.  We did not include binaries brighter 
than $m_{\rm F814W}=16.0$ in order to avoid the contamination from 
single MS and SGB stars with large photometric errors.  Moreover, 
we excluded binaries where the 2G star has $m_{\rm F814W}>18.5$ because 
we do not have any information on the colors of the fiducial lines 
at faint magnitudes and we would not predict the location in the CMD 
of the corresponding binaries.

The sample of selected binaries includes the 27 objects that are marked 
with orange triangles in Figure~\ref{fig:selBIN}.  The right panel of 
Figure~\ref{fig:selBIN} shows the $m_{\rm F814W}$ 
vs.\,$C_{\rm F275W,F336W,F438W}$ diagram of M\,4, where most of the 
selected binaries are located between the fiducial of single 1G stars 
and the fiducial of equal-mass 1G-1G binaries.  This diagram is used 
to derive the verticalized $m_{\rm F814W}$ 
vs.\,$\Delta$($C^{\rm bin}_{\rm F275W,F336W,F438W}$) diagram of the selected 
binaries that we plotted in the inset together with the corresponding
kernel-density and cumulative distributions of $\Delta(C^{\rm bin}_{\rm F275W,F336W,F438W})$.
 To derive the kernel-density distribution, which is used for illustration purposes only, 
 we adopted a Gaussian kernel with a fixed width that we derived with the rule of thumb by \citet[][]{silverman1986a}. 

The abscissa is calculated as:
\begin{equation}
\Delta(C^{\rm bin}_{\rm F275W,F336W,F438W})= [(X-X_{\rm fiducial}^{\rm 1G-1G})/(X_{\rm fiducial}^{\rm 1G}-X_{\rm fiducial}^{\rm 1G-1G})] 
\end{equation}
where $X$ is the $C_{\rm F275W,F336W,F438W}$ pseudo-color of the selected 
binaries, $X_{\rm fiducial}^{\rm 1G-1G}$ is the corresponding pseudo-color 
of the fiducial of equal-mass 1G-1G binaries and $X_{\rm fiducial}^{\rm 1G}$
is the  pseudo-color of the fiducial of single 1G stars.

\subsection{The incidence of binaries among stellar populations}
To infer the fraction of 1G-1G, 2G-2G and 1G-2G binaries with respect to 
the total number of binaries ($f_{\rm bin}^{\rm 1G-1G}$, 
$f_{\rm bin}^{\rm 2G-2G}$ and $f_{\rm bin}^{\rm 1G-2G}$), we compared 
the observations with a grid of simulated diagrams that are derived by using the ASs.  To do this, we 
defined a grid of values for $f_{\rm bin}^{\rm 1G-1G}$, 
$f_{\rm bin}^{\rm 2G-2G}$ and $f_{\rm bin}^{\rm 1G-2G}$ ranging from 
0.00 to 1.00 in steps of 0.01.  For each combination of 
$f_{\rm bin}^{\rm 1G-1G}$, $f_{\rm bin}^{\rm 2G-2G}$ and 
$f_{\rm bin}^{\rm 1G-2G}$, we compared the 
$\Delta$($C_{\rm F275W,F336W,F438W}$) kernel-density distribution 
of the simulated binaries with the observed distributions 
 and calculated the corresponding $\chi^{2}$.
 We assumed a flat mass-ratio distribution for simulated binaries as inferred by \citet{milone2012a} from observations of binaries in Galactic GCs. We also verified that the results remain unchanged  when we assume the two extreme mass ratio distributions used by \citet{sollima2007a} and \citet{milone2012a}. Specifically, we used the distribution obtained from random extractions from a \citet{demarchi2005a} initial mass function and the distribution measured by \citet{fisher2005a} and verified that the resulting values of $f_{\rm bin}^{\rm 1G-1G}$, $f_{\rm bin}^{\rm 2G-2G}$ and $f_{\rm bin}^{\rm 1G-2G}$ remain the same within 0.03.

\begin{centering} 
\begin{figure*} 
  \includegraphics[width=8.325cm,trim={0cm 5.cm 0.2cm 4.cm},clip]{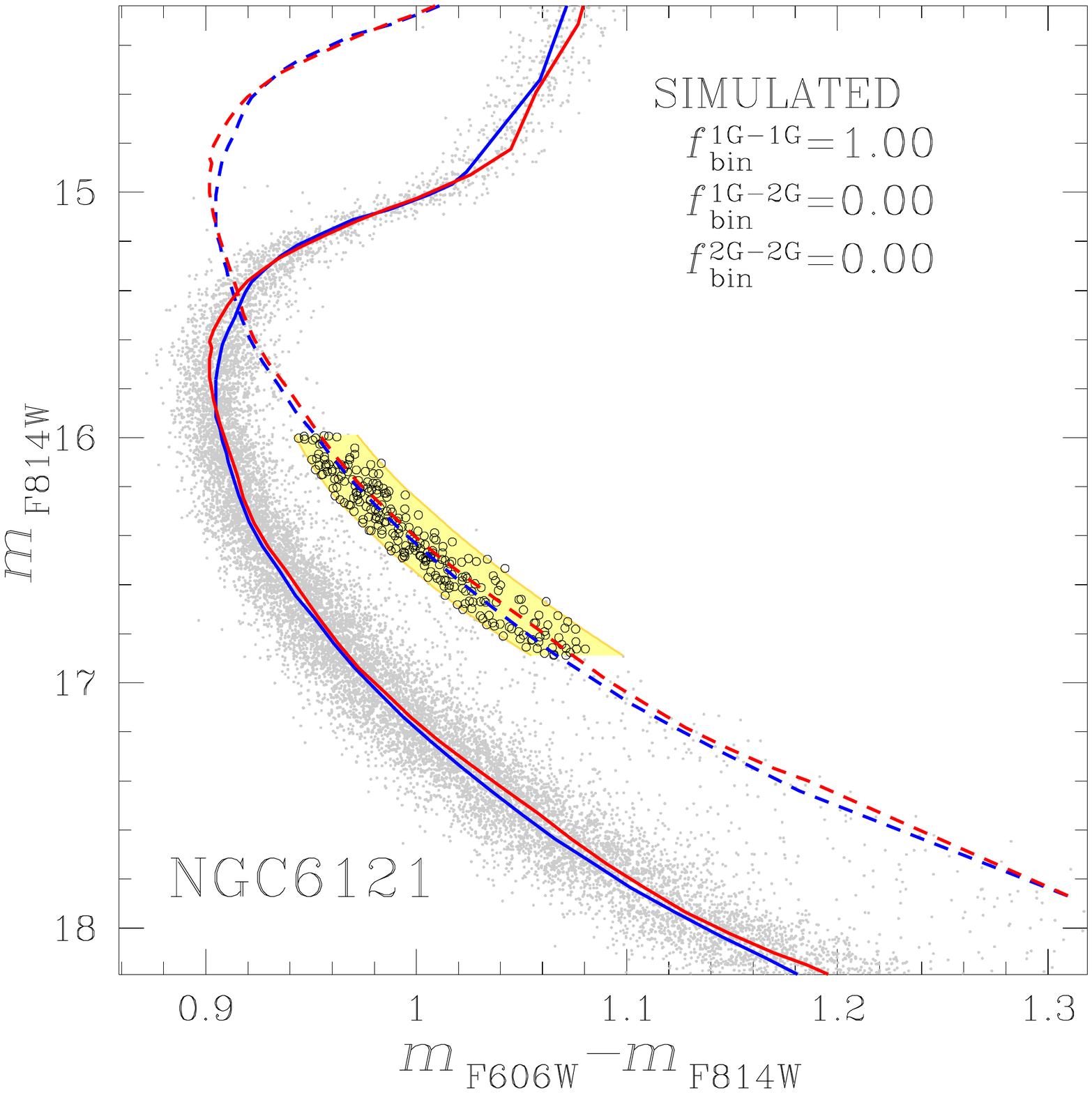}
  \includegraphics[width=8.25cm,trim={0.8cm 5.cm 0.2cm 2.7cm},clip]{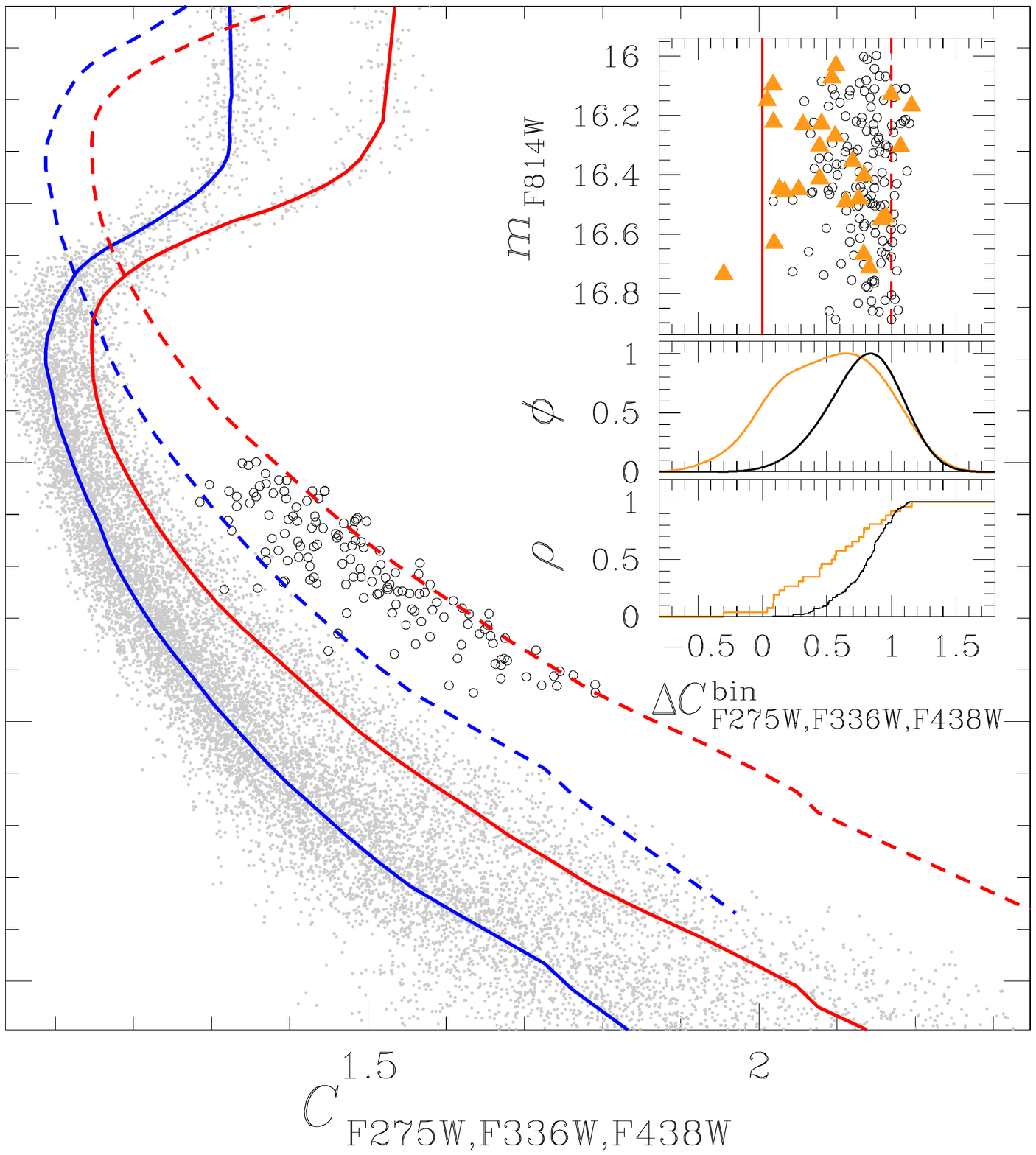}
  \includegraphics[height=8.25cm,trim={0cm 5.cm 0.8cm 2.7cm},clip]{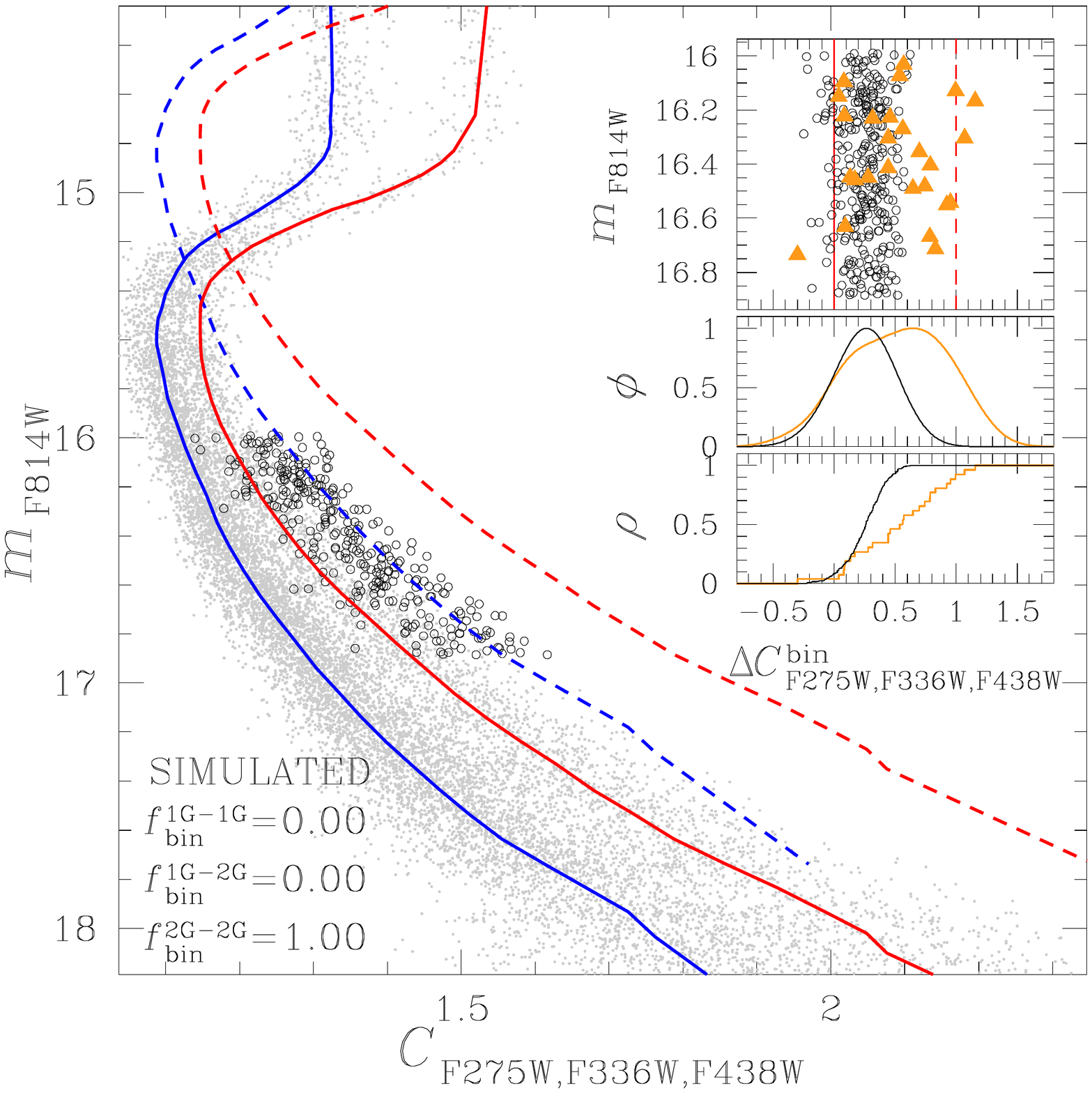}
  \includegraphics[height=8.25cm,trim={0.8cm 5.cm 0.2cm 2.7cm},clip]{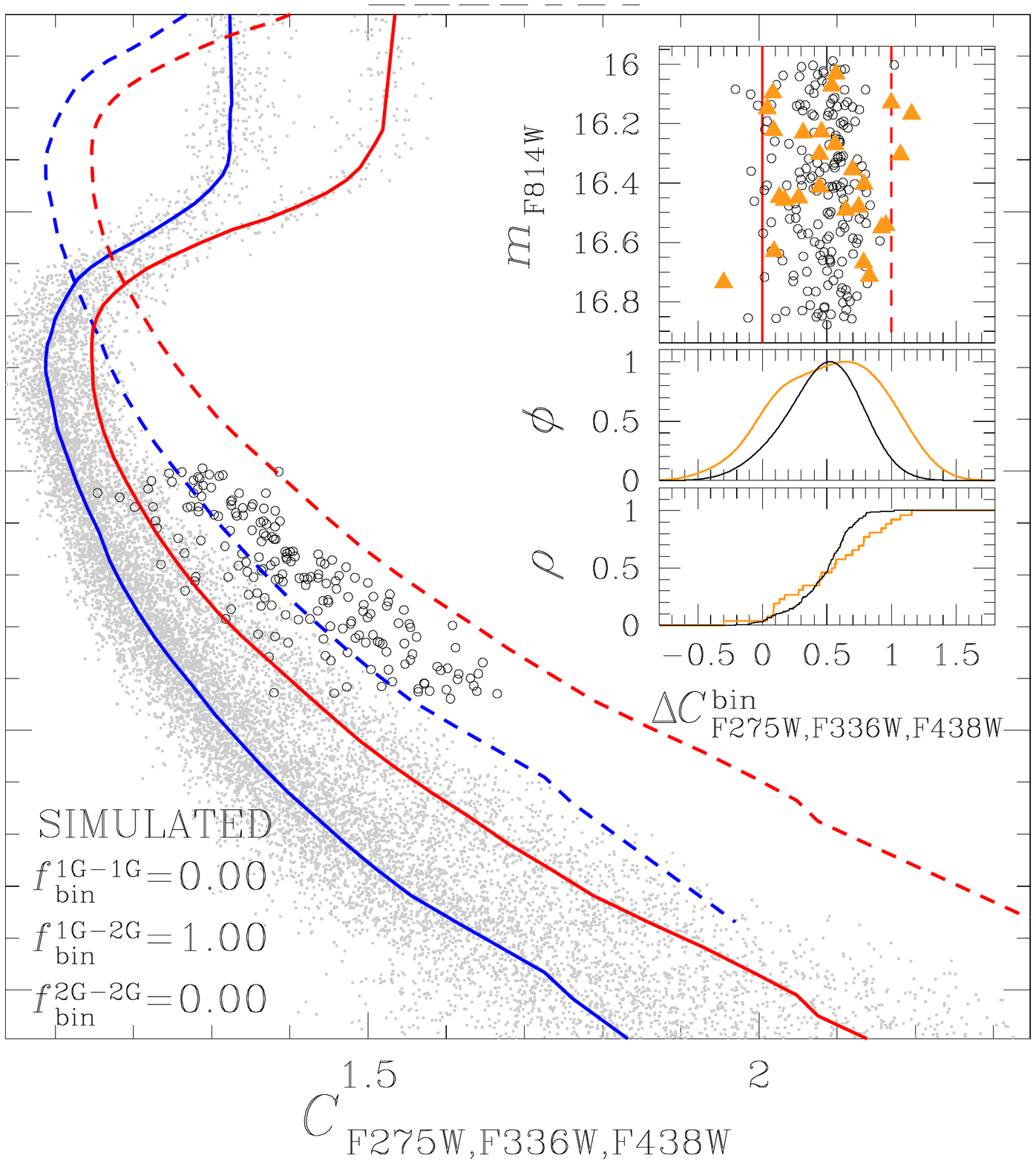}
  \caption{Upper panels show the simulated $m_{\rm F814W}$ vs.\,$m_{\rm F606W}-m_{\rm F814W}$  CMD (left) and $m_{\rm F814W}$ 
           vs.\,$C_{\rm F275W,F336W,F438W}$ diagram (right) where 
           we assumed that all the binaries are formed by pairs of 1G 
           stars. The yellow region includes all the selected binaries. 
           The inset compares the  $m_{\rm F814W}$ 
           vs.\,$\Delta C^{\rm bin}_{\rm F275W,F336W,F438W}$ diagram for the 
           selected sample of simulated (black circles) and observed 
           (orange triangles) binaries and the corresponding 
           $\Delta C^{\rm bin}_{\rm F275W,F336W,F438W}$ kernel distributions and cumulative distributions.
           In the bottom panels, we assumed 
           that all the binaries are formed by pairs of 2G stars (left) 
           and by pairs of 1G and 2G stars (right).  Simulated stars are colored gray, while the selected simulated binaries are marked with black circles.}
  \label{fig:simu1} 
\end{figure*} 
\end{centering}

As an example, we show in the upper panels of Figure~\ref{fig:simu1} the simulated  $m_{\rm F814W}$ vs.\,$m_{\rm F606W}-m_{\rm F814W}$ and $m_{\rm F814W}$ vs.\,$C_{\rm F275W,F336W,F438W}$ diagrams that  correspond to $f_{\rm bin}^{\rm 1G-1G}=1.00$, $f_{\rm bin}^{\rm 2G-2G}=0.00$  and $f_{\rm bin}^{\rm 1G-2G}=0.00$.  The yellow-shaded region defined in Figure~\ref{fig:selBIN} is used to identify the simulated stars that 
we compared with the sample of observed binaries. The selected simulated 
stars are marked with black circles in Figure~\ref{fig:simu1}.  
The $m_{\rm F814W}$ vs.\,$C_{\rm F275W,F336W,F438W}$ diagram shown 
in the right panel of Figure~\ref{fig:simu1} is used to derive the 
verticalized $m_{\rm F814W}$ vs.\,$\Delta$($C_{\rm F275W,F336W,F438W}$) 
diagram plotted in the inset, where we also compare the normalized cumulative distribution 
 and the kernel-density  distribution of the stars selected in the simulated diagrams (black lines) with the corresponding distributions derived in 
Figure~\ref{fig:selBIN} for the observed binaries (orange lines).  

The lower panels of Figure~\ref{fig:simu1} shows the simulated $m_{\rm F814W}$  vs.\,$C_{\rm F275W,F336W,F438W}$ diagrams for different choices of 
$f_{\rm bin}^{\rm 1G-1G}$, $f_{\rm bin}^{\rm 2G-2G}$ and 
$f_{\rm bin}^{\rm 1G-2G}$.  In the lower-left panel we assumed that 
all the binary systems  are composed of 2G-2G pairs while all the binaries in the right lower-panel include both 1G and 2G stars. 
In both cases, we obtain a poor match to the observations, as shown by 
the verticalized diagrams and by the corresponding  cumulative and kernel-density 
distributions plotted in the insets.

\begin{centering} 
\begin{figure} 
  \includegraphics[width=9.00cm,trim={0cm 5.cm 0.2cm 2.cm},clip]{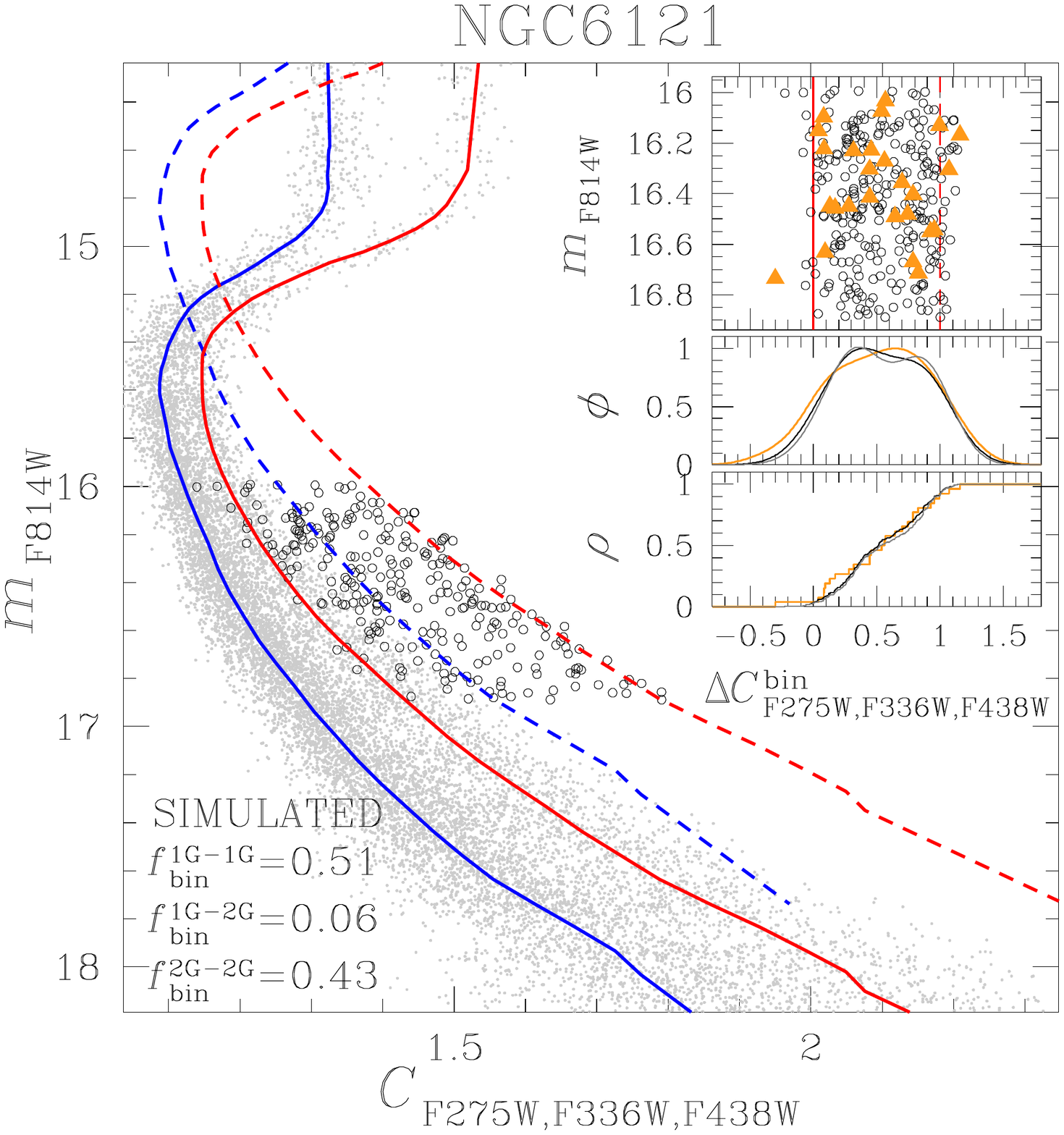}
  \caption{Simulated $m_{\rm F814W}$ vs.\,$C_{\rm F275W,F336W,F438W}$ diagram
    of NGC\,6121. 
     The assumed fractions of 1G-1G, 1G-2G, and 2G-2G
     binaries are quoted in the figure and correspond to the simulation that
     provides the best match with the observations.
      Gray and black colors indicate the simulated stars and the selected simulated binaries, respectively.
 The inset compares the $m_{\rm F814W}$ vs.\,$\Delta C^{\rm bin}_{\rm F275W,F336W,F438W}$ and the corresponding $\Delta C^{\rm bin}_{\rm F275W,F336W,F438W}$ kernel-density distribution  and cumulative distribution for simulated (black) and observed binaries (orange).  Gray lines correspond to the distributions of $\Delta C^{\rm bin}_{\rm F275W,F336W,F438W}$ for the simulation with no mixed binaries that provides the best match with the data. }
 \label{fig:simu2} 
\end{figure} 
\end{centering}

Finally, in Figure~\ref{fig:simu2} we show the simulated diagrams that provide the best match with the observations,  which is derived as the minimum difference between the corresponding normalized cumulative distributions as plotted in the bottom panel of the inset. 
 The best-fit corresponds to $f_{\rm bin}^{\rm 1G-1G}=0.51$, $f_{\rm bin}^{\rm 1G-2G}=0.06$ and $f_{\rm bin}^{\rm 2G-2G}=0.43$.   For completeness, we compare in the middle panel of the inset the kernel-density distribution of $\Delta C^{\rm bin}_{\rm F275W,F336W,F438W}$ for the observed and the simulated binaries.

The uncertainties associated with these values are calculated with a bootstrap analysis based on 30,000 samples created by a random sampling with replacement of the observed binary stars. For each extraction we derived the fraction of 1G-1G, 1G-2G and 2G-2G binaries by using the procedure described above.

The obtained random mean scatter of the 30,000 determinations of the values
of $f_{\rm bin}^{\rm 1G-1G}$, $f_{\rm bin}^{\rm 1G-2G}$ and 
$f_{\rm bin}^{\rm 2G-2G}$ are 0.11,  0.04, and 0.10, respectively, and 
are considered as the best estimates of the corresponding uncertainties.

To investigate whether the inferred results are reliable or not, we used 
ASs to generate 30,000 mock CMDs that host the same fraction of 1G-1G, 
1G-2G, and 2G-2G binaries that we inferred from the observations. 
We selected 27 stars from each simulation that are located in the same 
region of the $m_{\rm F814W}$ vs.\,$m_{\rm F606W}-m_{\rm F814W}$ CMD 
defined in Figure~\ref{fig:selBIN} to select the sample of binaries in 
the observed CMD.

We calculated the values of $f_{\rm bin}^{\rm 1G-1G}$, 
$f_{\rm bin}^{\rm 1G-2G}$ and $f_{\rm bin}^{\rm 2G-2G}$ in each simulation 
by using the same procedure described above for real stars.  The average 
values of 1G-1G, 1G-2G, and 2G-2G binary fractions that we obtained from 
the 30,000 simulated CMDs are identical to the values that we inferred 
from the observations,  while the uncertainties associated to $f_{\rm bin}^{\rm 1G-1G}$, $f_{\rm bin}^{\rm 1G-2G}$ and $f_{\rm bin}^{\rm 2G-2G}$ are slightly smaller and correspond to 0.09, 0.03, and 0.09, respectively. These results ensure that the adopted procedure does not introduce any significant systematic error. 

 Results suggest that about 6\% of the studied binaries of NGC\,6121 are formed by pairs of 1G and 2G stars, but this result is significant at $\sim 1.5 \sigma$-level only.
To better understand how significant is the detection of the mixed 1G-2G population
 we used the procedure described above to derive the best fit simulation containing only 1G-1G and 2G-2G binaries. The resulting cumulative and kernel-density distributions of $\Delta C^{\rm bin}_{\rm F275W,F336W,F438W}$ are represented with gray lines in the inset of Figure~\ref{fig:simu2} and correspond to the simulation composed of 0.52$\pm$0.12 and 0.48$\pm$0.12 of 1G-1G amd 2G-2G binaries. 
 The Kolmogorov-Smirnov (KS) test provides a probability p=57\% that the binaries from best-fit simulation and the observed binaries come from the same parent distribution. The corresponding probability inferred from the comparison of the observations with the best-fit model that accounts for mixed binaries is p=92\% and seems to corroborate the conclusion that NGC\,6121 hosts a small fraction of mixed binaries. 

 

\begin{table*}
 \caption{This table lists for each cluster the fraction of 1G stars with respected to the total number of MS stars ($N_{\rm 1G}/N_{\rm TOT}$), the number of analyzed binaries ($N_{\rm bin}$), the fractions of 1G-1G and 2G-2G binaries ($f_{\rm bin}^{\rm 1G-1G}$, $f_{\rm bin}^{\rm 1G-2G}$, $f_{\rm bin}^{\rm 2G-2G}$ ),
  the 
 ratio of the incidence of 1G-1G binaries among 1G stars to the incidence of 2G-2G binaries among 2G stars,  
 $f_{\rm b,1G}/f_{\rm b,2G}$, and the fraction of primordial binaries $f_{\rm pri}$.}
 \begin{tabular}{l c cccc cc}
   \hline
   \hline
 ID  & $N_{\rm 1G}/N_{\rm TOT}$  & $N_{\rm bin}$ & $f_{\rm bin}^{\rm 1G-1G}$ & $f_{\rm bin}^{\rm 1G-2G}$ & $f_{\rm bin}^{\rm 2G-2G}$  & $f_{\rm b,1G}/f_{\rm b,2G}$ & $f_{\rm pri}$ \\
\hline
NGC\,288  & 0.56$\pm$0.01 & 95 & 0.46$\pm$0.08 & 0.14$\pm$0.07 & 0.40$\pm$0.08 & 1.0$\pm$0.3 & 0.72$\pm$0.15 \\
NGC\,6121 & 0.29$\pm$0.01 & 27 & 0.51$\pm$0.10 & 0.06$\pm$0.04 & 0.43$\pm$0.10 & 3.1$\pm$0.9 & 0.85$\pm$0.10 \\
NGC\,6352 & 0.50$\pm$0.01 & 65 & 0.24$\pm$0.10 & 0.48$\pm$0.09 & 0.28$\pm$0.07 & 0.9$\pm$0.4 & 0.00$\pm$0.18\\
NGC\,6362 & 0.55$\pm$0.01 & 74 & 0.47$\pm$0.07 & 0.00$\pm$0.03 & 0.51$\pm$0.07 & 0.7$\pm$0.2 & 1.00$\pm$0.06\\
NGC\,6838 & 0.63$\pm$0.01 & 46 & 0.46$\pm$0.13 & 0.27$\pm$0.13 & 0.27$\pm$0.09 & 1.2$\pm$0.4 & 0.42$\pm$0.28\\
 \hline\hline
 \end{tabular}\\
 \label{tab:res}
 \end{table*}

\begin{centering} 
\begin{figure*} 
  \includegraphics[height=6.75cm,trim={0cm 5.cm 0.2cm 5.cm},clip]{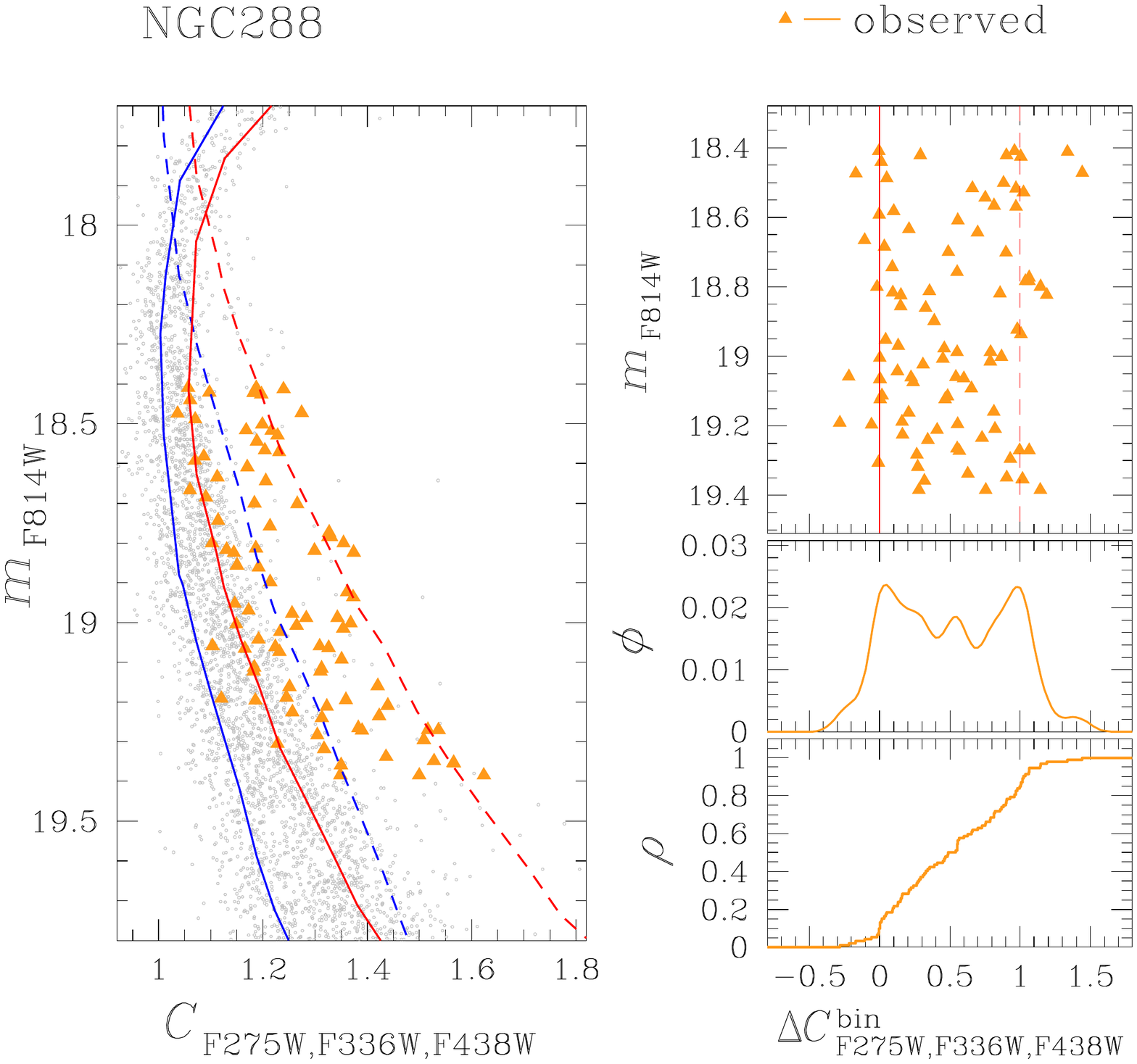}
  \includegraphics[height=6.75cm,trim={10.2cm 5.cm 0.2cm 5.cm},clip]{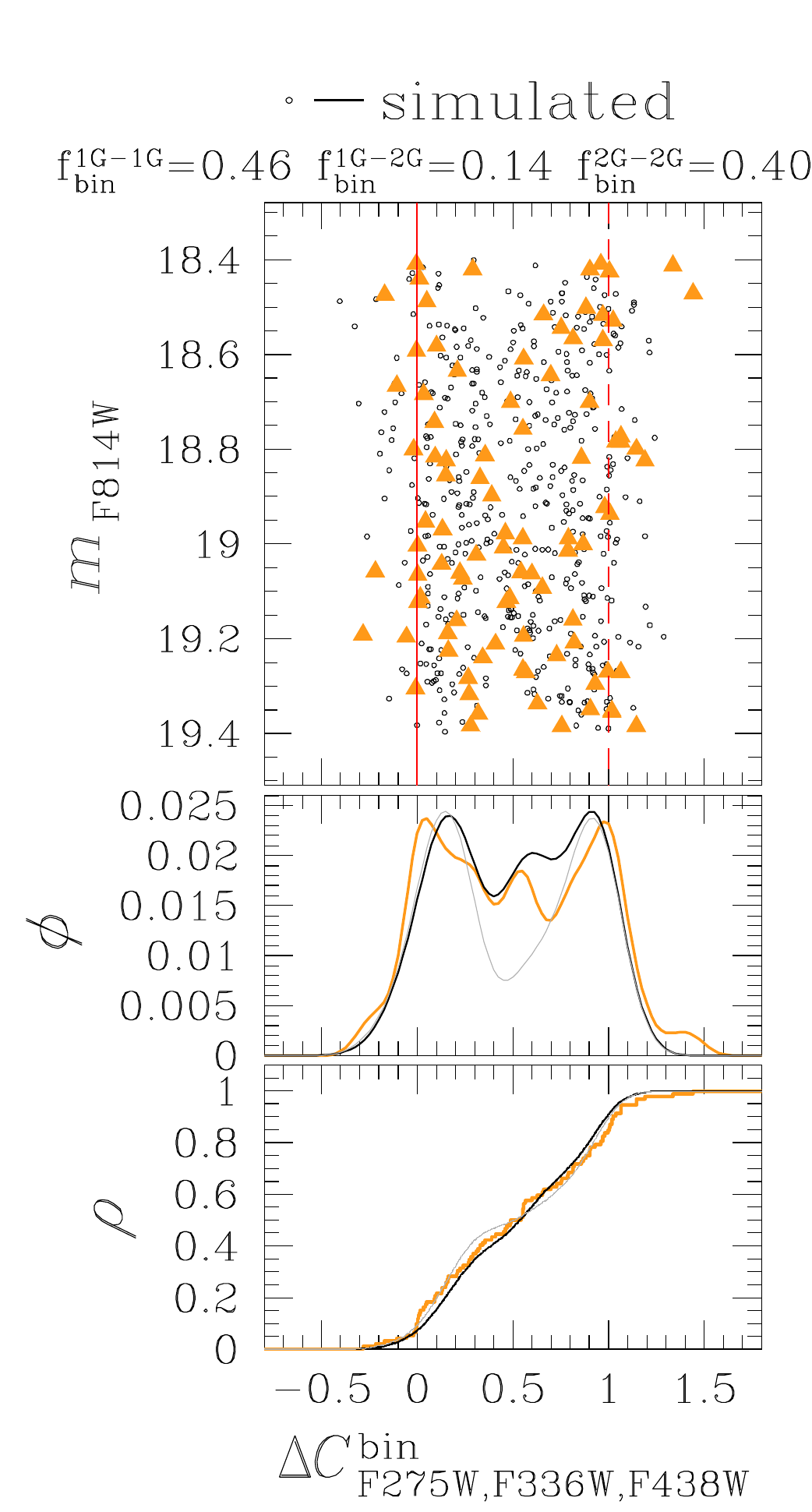}
  \includegraphics[height=6.75cm,trim={0cm 5.cm 0.2cm 5.cm},clip]{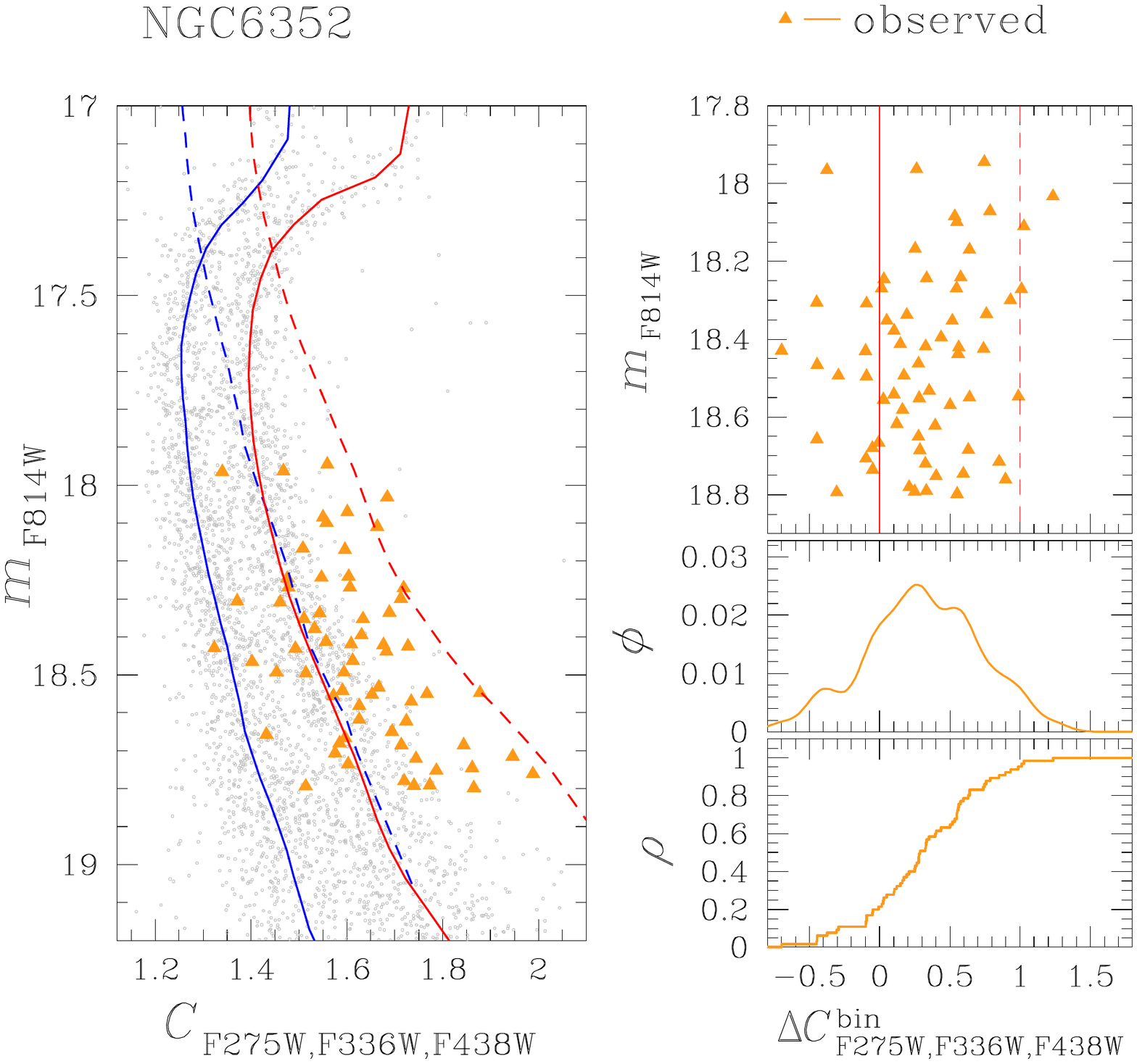}
  \includegraphics[height=6.75cm,trim={10.2cm 5.cm 0.2cm 5.cm},clip]{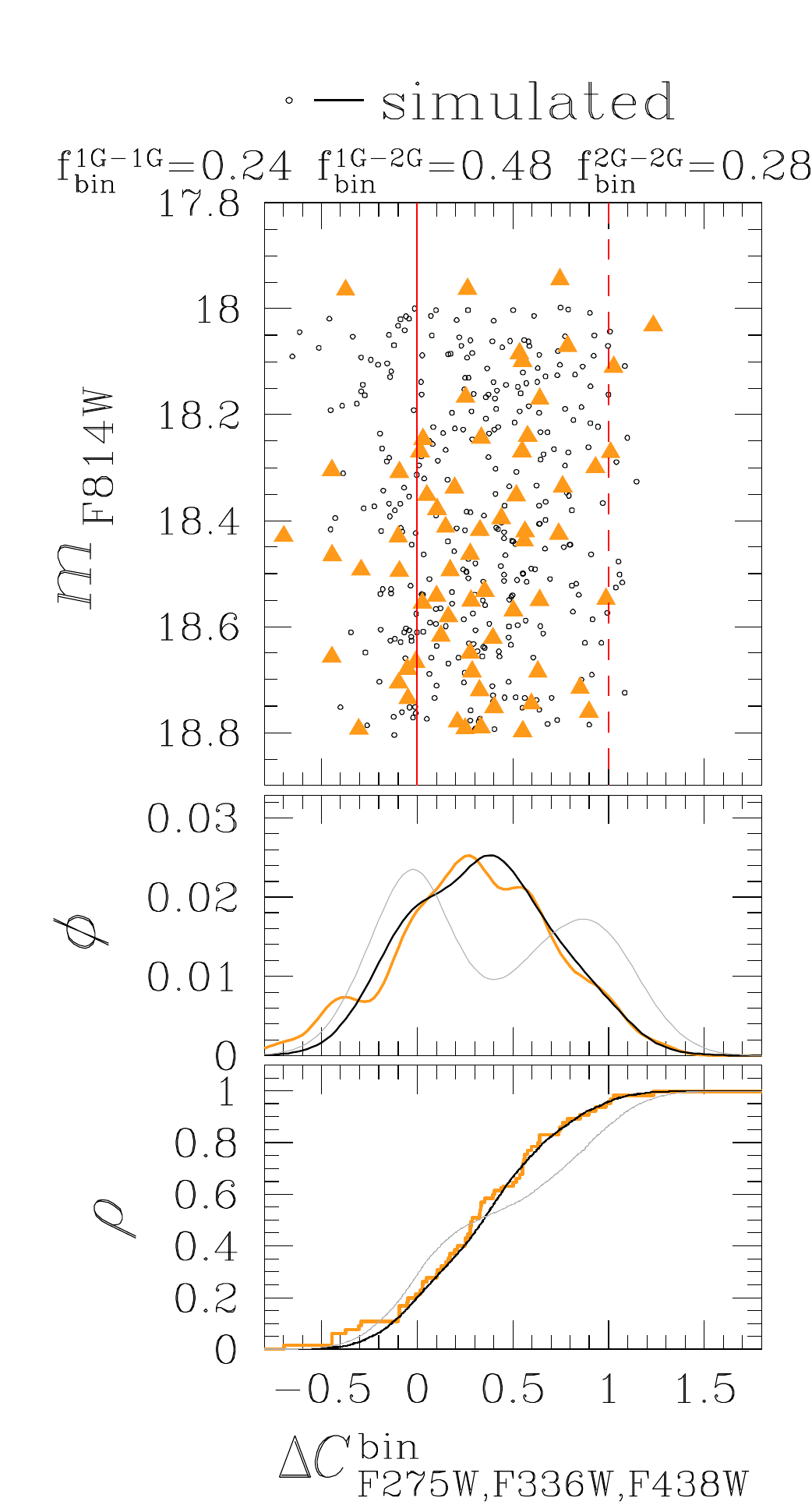}
  \caption{ $m_{\rm F814W}$ vs.\,$C_{\rm F275W,F336W,F438W}$ diagrams zoomed in the upper MS (left).  Verticalized $m_{\rm F814W}$ vs.\,$\Delta C^{\rm bin}_{\rm F275W,F336W,F438W}$ for the selected binaries and corresponding kernel-density distribution (middle and right panels).  Red and blue continuous lines are the fiducials of 1G and 2G stars, respectively, while the dashed lines with the same color are the corresponding fiducials for equal-mass binaries. Orange triangles mark the selected observed binaries, whose $\Delta C^{\rm bin}_{\rm F275W,F336W,F438W}$ kernel-density distribution and cumulative distribution is represented with orange lines. Black dots and black lines refer to the simulated binaries. 
   The gray lines correspond to the distributions of $\Delta C^{\rm bin}_{\rm F275W,F336W,F438W}$ for the simulations with no mixed binaries that provides the best match with the data.  The fraction of 1G-1G, 1G-2G and 2G-2G simulated binaries are quoted in the right panels.} 
 \label{fig:resall1} 
\end{figure*} 
\end{centering}

\begin{centering} 
\begin{figure*} 
  \includegraphics[height=6.75cm,trim={0cm 5.cm 0.2cm 5.cm},clip]{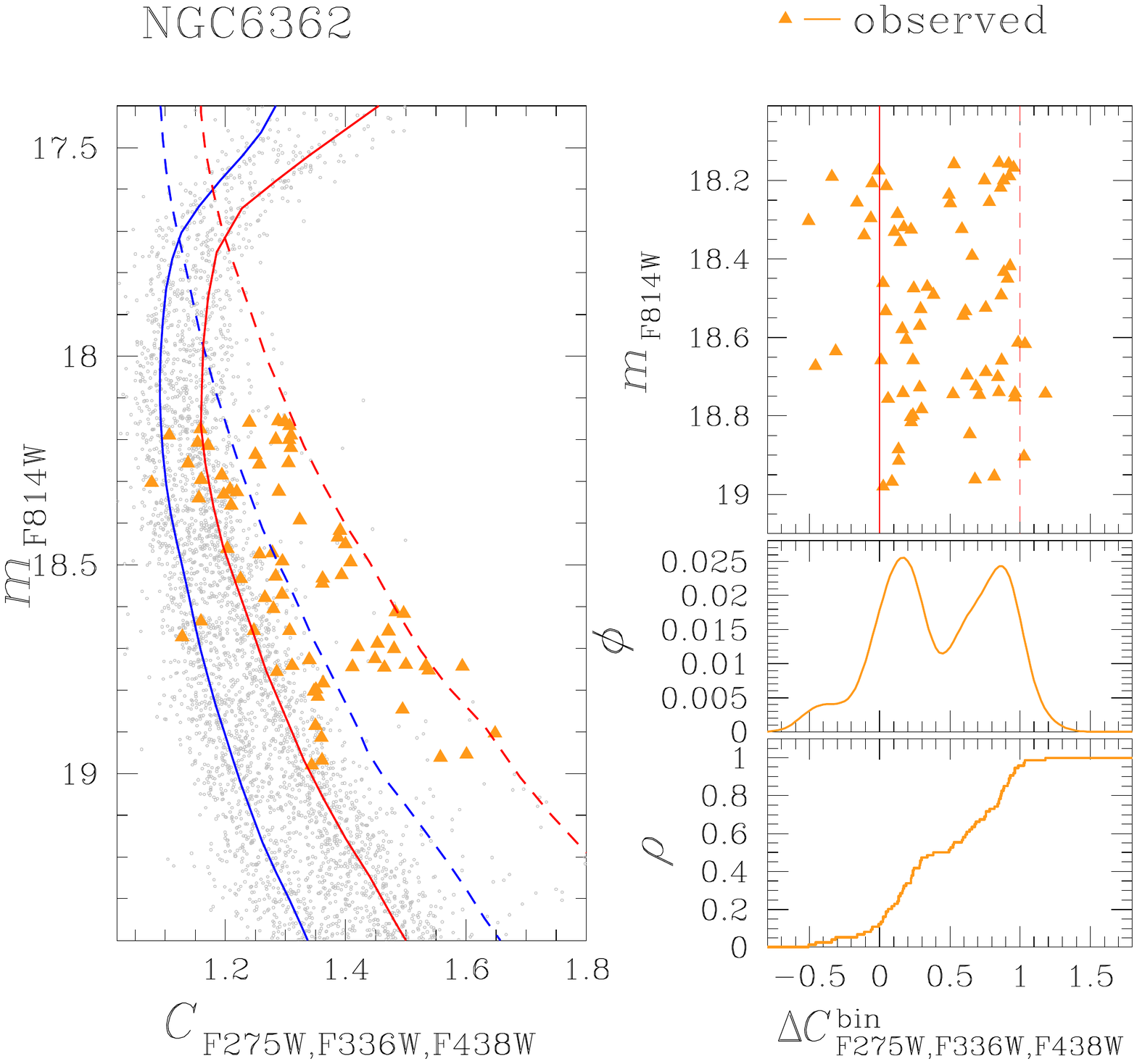}
  \includegraphics[height=6.75cm,trim={10.2cm 5.cm 0.2cm 5.cm},clip]{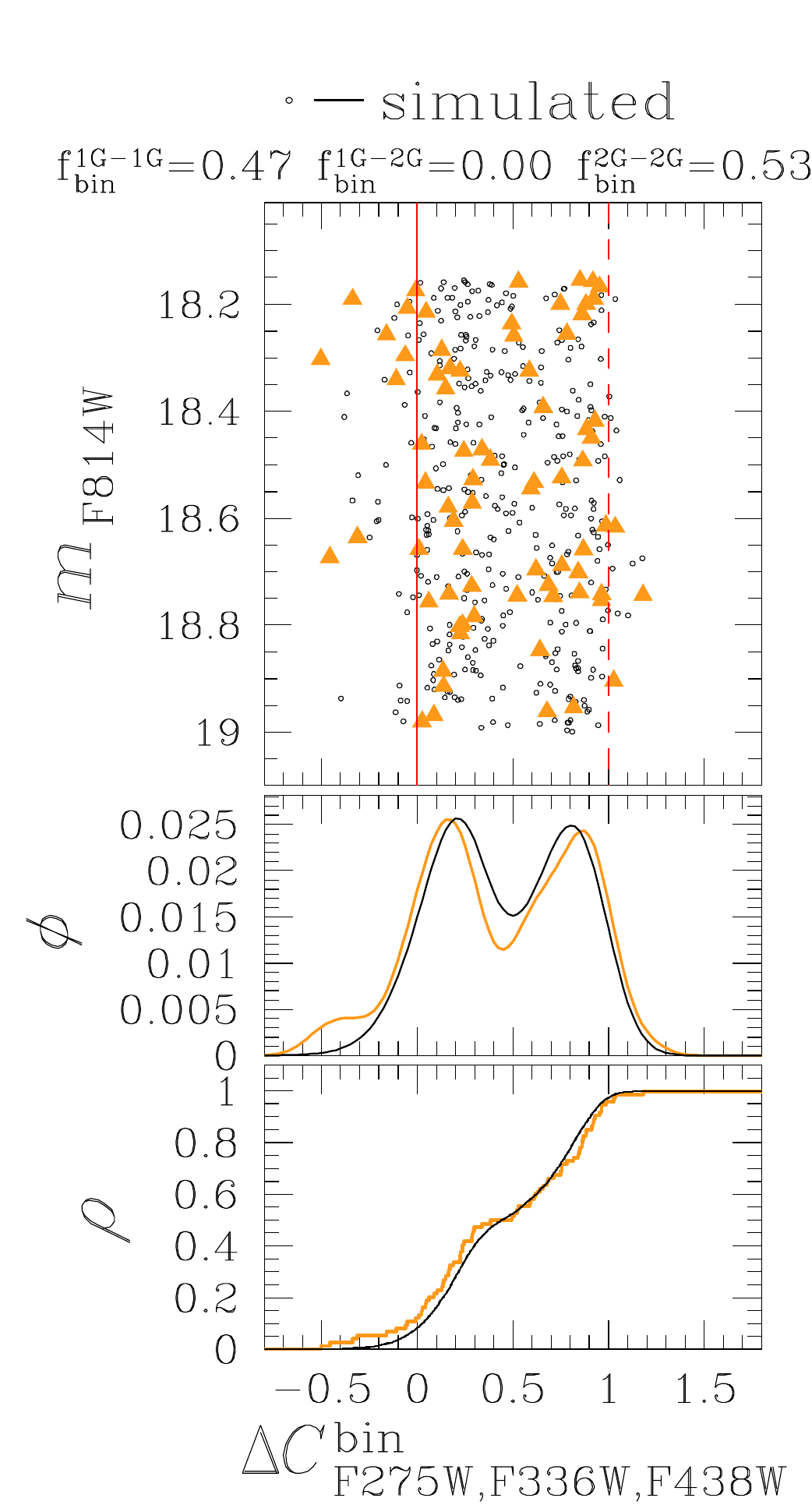}
  \includegraphics[height=6.75cm,trim={0cm 5.cm 0.2cm 5.cm},clip]{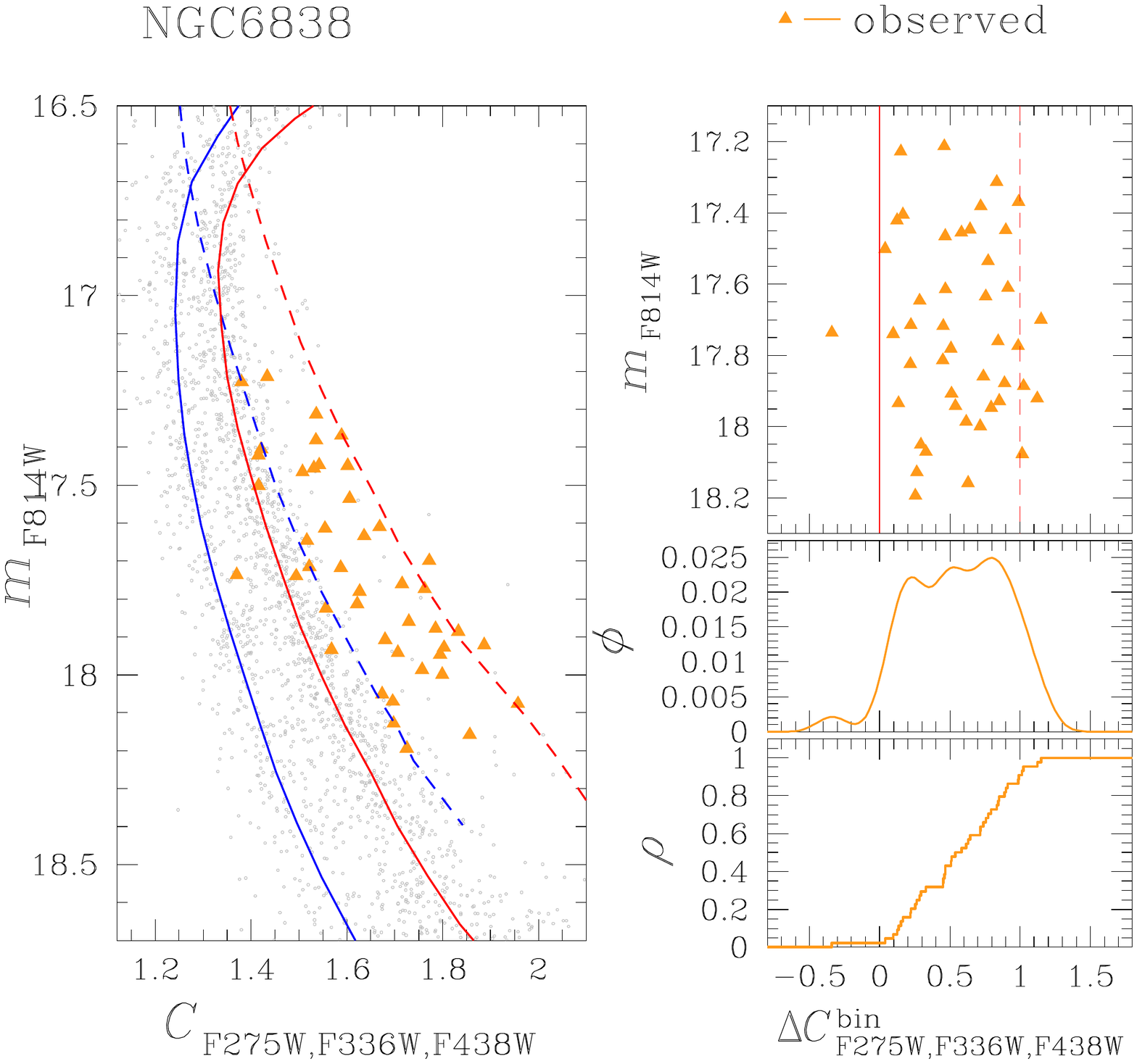}
  \includegraphics[height=6.75cm,trim={10.2cm 5.cm 0.2cm 5.cm},clip]{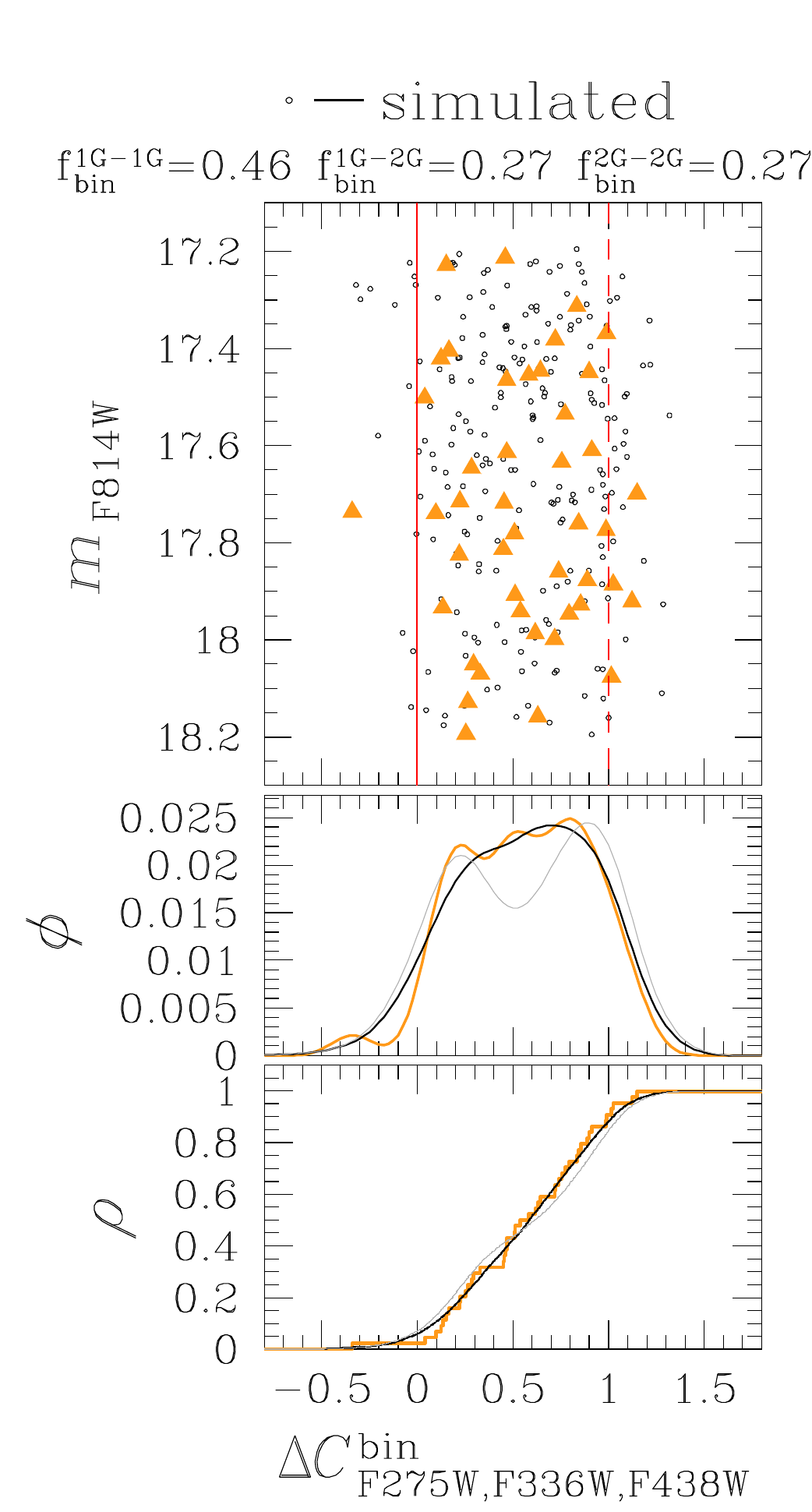}
  \caption{As in Figure~\ref{fig:resall1} but for NGC\,6362 and NGC\,6838.} 
 \label{fig:resall2} 
\end{figure*} 
\end{centering}

The procedure described above for NGC\,6121 was extended to the other clusters and the main results are shown in Figures~\ref{fig:resall1} and \ref{fig:resall2} and summarized in Table~\ref{tab:res}.
Left panels of  these figures are zoom in of the $m_{\rm F814W}$ vs.\,$C_{\rm F275W,F336W,F438W}$ diagrams around the upper MS, while middle panels show $m_{\rm F814W}$ against $\Delta C^{\rm bin}_{\rm F275W,F336W,F438W}$ for the sample of selected binaries and the corresponding  cumulative and kernel-density distributions.

The $\Delta C^{\rm bin}_{\rm F275W,F336W,F438W}$ distributions of binaries in NGC\,288
and NGC\,6362 are clearly bimodal with two main groups of stars with $\Delta C^{\rm bin}_{\rm F275W,F336W,F438W} \sim 0.0-0.1$ and $\sim 0.8-1.0$. In contrast, a single peak with intermediate values of $\Delta C^{\rm bin}_{\rm F275W,F336W,F438W} \sim 0.3$ is present in NGC\,6352, while the binaries of NGC\,6838 exhibit a broad distribution.

Right panels of Figures~\ref{fig:resall1}  and \ref{fig:resall2} show the distribution of binaries from the simulated diagrams that provide the best match with the observations and are obtained from the comparison of the corresponding normalized cumulative distributions.
Although the results are inferred from a large sample of simulated binaries as described above, for clarity, the number of binaries that we plotted in each figure as black dots is equal to five times the number of observed binaries.
We find that, similarly to NGC\,6121, both NGC\,288 and NGC\,6362 host small fractions of 1G-2G stars, and comparable fractions of 1G-1G and 2G-2G binaries. This fact explains the bimodal $\Delta C^{\rm bin}_{\rm F275W,F336W,F438W}$ distributions of the observed binaries.
In the case of NGC\,6352, we find that about half of the studied binary systems are 1G-2G pairs, while the fraction of 1G-1G and 2G-2G binaries are similar. The predominance of mixed binaries is responsible for the single peak of the kernel-density distribution with intermediate values of $\Delta C^{\rm bin}_{\rm F275W,F336W,F438W}$. NGC\,6838 hosts a large fraction of 1G-2G binaries ($f_{\rm bin}^{\rm 1G-2G}\sim$0.27) and a similar fraction of 2G-2G pairs.

To estimate the incidence of 1G-1G binaries among 1G star with respect to incidence of 2G-2G binaries among 2G stars we calculate the quantity:
$f_{\rm b,1G}/f_{\rm b,2G}=(f_{\rm bin}^{\rm 1G-1G}/N_{\rm 1G})/(f_{\rm bin}^{\rm 2G-2G}/N_{\rm 2G})$, where $N_{\rm 1G}$ and where $N_{\rm 2G}$, are the numbers of analyzed 1G and 2G MS stars. Results are listed in Table~\ref{tab:res}. In M\,4 we find that the fraction of 1G-1G binary pairs among 
1G stars is $\sim 3$ times higher than the fraction of 2G-2G binaries among 2G stars and the difference is significant at $\sim$3-$\sigma$ level. In the other clusters $f_{\rm b,1G}/f_{\rm b,2G}$ is consistent with one.

To further investigate the significance of the detection of mixed binaries in NGC\,288, NGC\,6352 and NGC\,6838 we derived the simulation with $f_{\rm bin}^{\rm 1G-2G}=0$ that best reproduce the observations.

Specifically, the results listed in Table~\ref{tab:res} indicate that about half of the binaries of NGC\,6352 are formed by pairs of 1G-2G stars and the detection of mixed binaries is significant at $\sim 4 \sigma$ level.
The KS test indicates that the binaries of the best-fit simulation obtained for NGC\,6352 has a probability higher than 0.99 to come from the same parent distribution of the observed binaries. In contrast, the corresponding probability for best-fit simulation formed by 1G-1G and 2G-2G binaries alone is 0.00. This fact confirms the high significance of detection of mixed binaries in this GC. 

In NGC\,288 and NGC\,6838 the best-fit simulations with no mixed binaries that provide KS probabilities of 0.31 and 0.11, respectively, which are lower than the corresponding probabilities of 0.90 and 0.98, respectively, derived from the best-fit models that account for 1G-2G binaries  although still statistically compatible with observations. These findings are in line with the results of Table~\ref{tab:res}, where we estimate that the detection of mixed binaries in each cluster is significant at $\sim 2$-$\sigma$ level.

\subsection{Primordial and dynamically-formed binaries}

\begin{centering} 
\begin{figure} 
  \includegraphics[width=8.5cm,trim={0.0cm 0.0cm 6.2cm 16.0cm},clip]{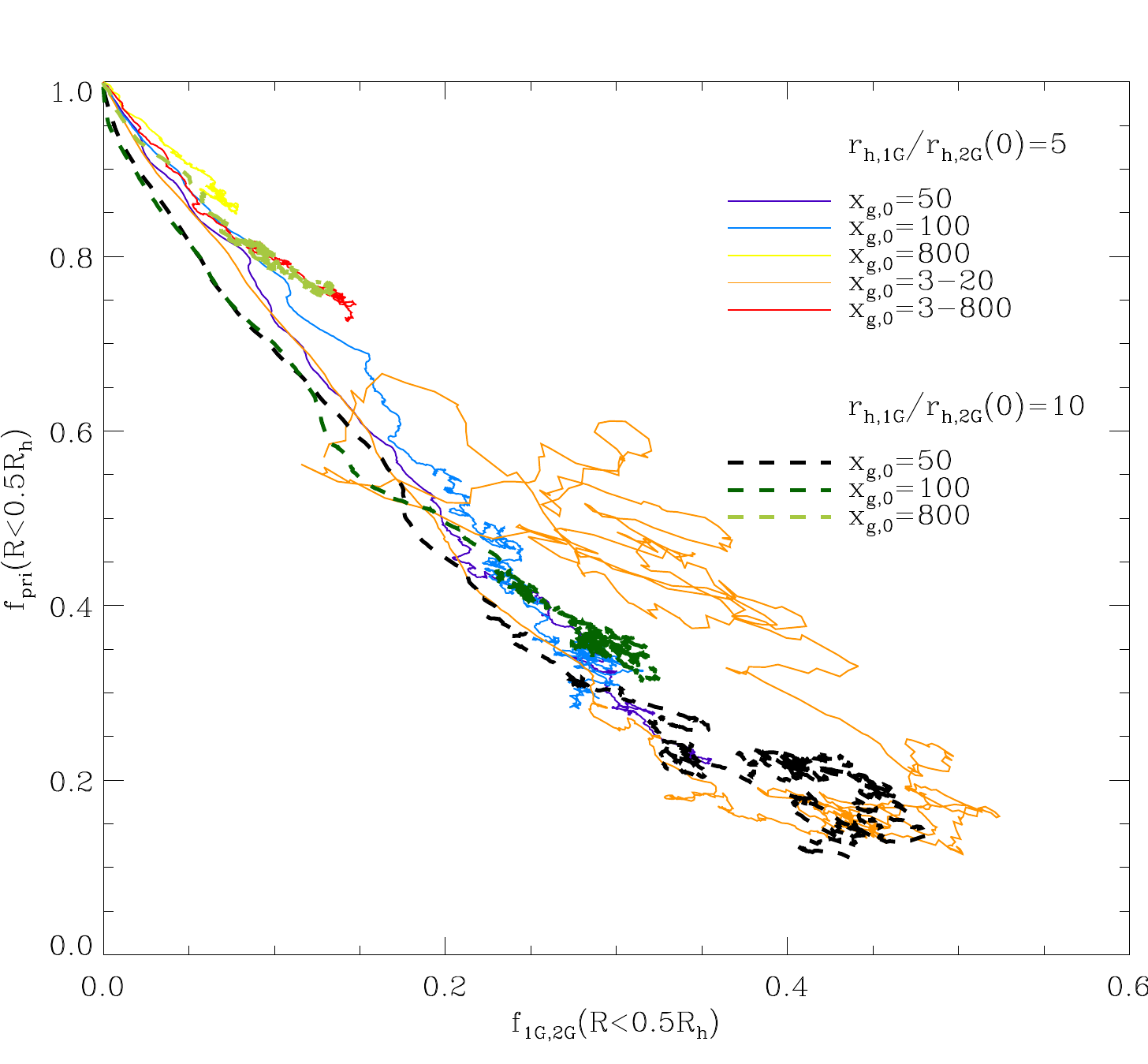}
  \caption{Evolution of the simulated fraction of primordial binaries against the fraction of of mixed binaries (see text for details on the N-body simulations).}
 \label{fig:primordial} 
\end{figure} 
\end{centering} 

Present-day binaries in GCs include primordial binaries, which have origin from the same gas cloud and include only 1G-1G or 2G-2G binaries, and binaries formed during the cluster's dynamical evolution from capture and/or exchange events which can pair stars of different generations and produce some mixed 1G-2G binaries \citep{hong2015a, hong2016a}.

Here, we have used the results of a set of N-body simulations following the evolution of binaries in multiple-population clusters \citep{hong2015a, hong2016a} to establish a link between the fraction of mixed binaries and the fraction of observed binaries belonging to the primordial binary population. To further illustrate this link we have also built a binary population from a Monte Carlo sampling procedure from the observed fraction of 1G and 2G stars.

In Figure~\ref{fig:primordial}, we show the evolution of the fraction of primordial binaries in the total population of binaries  versus the fraction of mixed 1G-2G binaries from our N-body simulations (further details on the simulations are discussed later in Section 5).
This figure clearly illustrates the dynamical information encoded in the fraction of mixed binaries: as a cluster evolves, its binary population is affected by stellar encounters, the fraction of mixed binaries increases, and the fraction of primordial binaries in the binary population declines. Some primordial binaries are disrupted, some are ejected, and some undergo exchange encounters resulting in binaries with components different from those in the primordial binary.
Although the simulations are still idealized and not meant to provide detailed models for the observed clusters, the observed values of the fraction of mixed binaries reported in Table~1 and the data shown in Figure~\ref{fig:primordial} can be used to calculate an approximate estimate of the fraction of the observed binaries belonging to the primordial binary population.

In order to further explore the link between the fraction of mixed binaries and the fraction of primordial binaries in the current binary population we have also carried out  101 Monte Carlo samplings of 100,000 MS stars. 
In each simulation, $i$, we included a fraction of primordial binaries $f_{\rm bin, i}^{\rm pri, simu}$=$i$/100, where $i$ ranges from 0 to 100 in steps of 1. The remaining simulated stars, which comprise the observed fractions of 1G and 2G stars, are randomly coupled. Clearly, this process generates pairs of 1G-1G, 1G-2G, and 2G-2G binaries. 

We indicate the resulting fraction of 1G-2G binaries with respect to the total number of binaries (including both primordial binaries, and binaries derived by random pair stars) as $f_{\rm bin, i}^{\rm 1G-2G, simu}$. The observed binaries with a primordial origin in each cluster, $f_{\rm pri}$, is provided by the simulation where $f_{\rm bin, i}^{\rm 1G-2G, simu}$ matches the observed fraction of mixed binaries.
Results are listed in Table\ref{tab:res}.
The estimates of the fraction of primordial binaries obtained from simulations are in general good agreement with those found with the Monte Carlo sampling procedure; in particular,
we find that the NGC\,288, NGC\,6121 and NGC\,6362 are dominated by primordial binaries, while NGC\,6352 is consistent with almost no primordial binaries. About half of the studied binaries of NGC\,6838 have primordial origins.  

We emphasize that these estimates are meant to provide a general approximate indication of the fraction of primordial binaries and, more in general, to illustrate the dynamical information contained in the population of mixed binaries. More realistic models would be necessary to use the observed fraction of mixed binaries to obtain accurate estimates of the primordial binary fraction.

\section{Discussion}\label{sec:discussion}
The present-day binary fractions of 1G and 2G stars provide a dynamical fingerprint of the formation and dynamical evolution of multiple populations in GCs.
According to various scenarios, 2G stars form in a dense environment in 
the innermost regions of a more extended 1G system \citep[e.g.\,][and references therein]{dercole2008a, calura2019a}.
 Analytic calculations combined with the results of $N$-body simulations 
of stellar populations in GCs show that, as a consequence of these initial differences between the spatial distributions of 1G and 2G stars, 2G binaries evolve and are disrupted at a significantly larger rate than 1G binaries and the 
present-day 2G population is expected to have a smaller {\it global} binary incidence than the 1G population \citep{vesperini2011a, hong2015a, hong2016a}.   
The evolution of the ratio of the 1G to the 2G binary incidence is driven by the initial differences between the structural properties of the 1G and the 2G populations and depends on the cluster's dynamical age as well as on the binary properties \citep[see e.g.][]{hong2015a, hong2016a, hong2019a}.

The complex interplay between binary evolution, disruption, and the evolution of the spatial distributions of 1G and 2G single and binary stars is expected to result into a radial variation of the 1G and 2G binary incidences that need to be taken into account in the interpretation of observational data that probe only a specific range of radial distances from the cluster's center and thus provide a measure of the {\it local} binary incidence and not the {\it global} one. 
This issue has been discussed in detail in \citet{hong2016a} (see, in particular, their Figures 11 and 12). Hong and collaborators found that the largest differences between the 1G and the 2G binary incidences are, in general, expected in the cluster's outer regions (see, for example, their Figure\,12 showing the time evolution of the ratio 
 $f_{\rm b, 1G}/f_{\rm b, 2G}$ estimated at projected distances between 0.5$R_{\rm h}$ and 2.5 $R_{\rm h}$ where $R_{\rm h}$ is the projected half-mass radius). 

\begin{centering} 
\begin{figure} 
  \includegraphics[width=8.5cm,trim={0cm 0cm 6.2cm 13.0cm},clip]{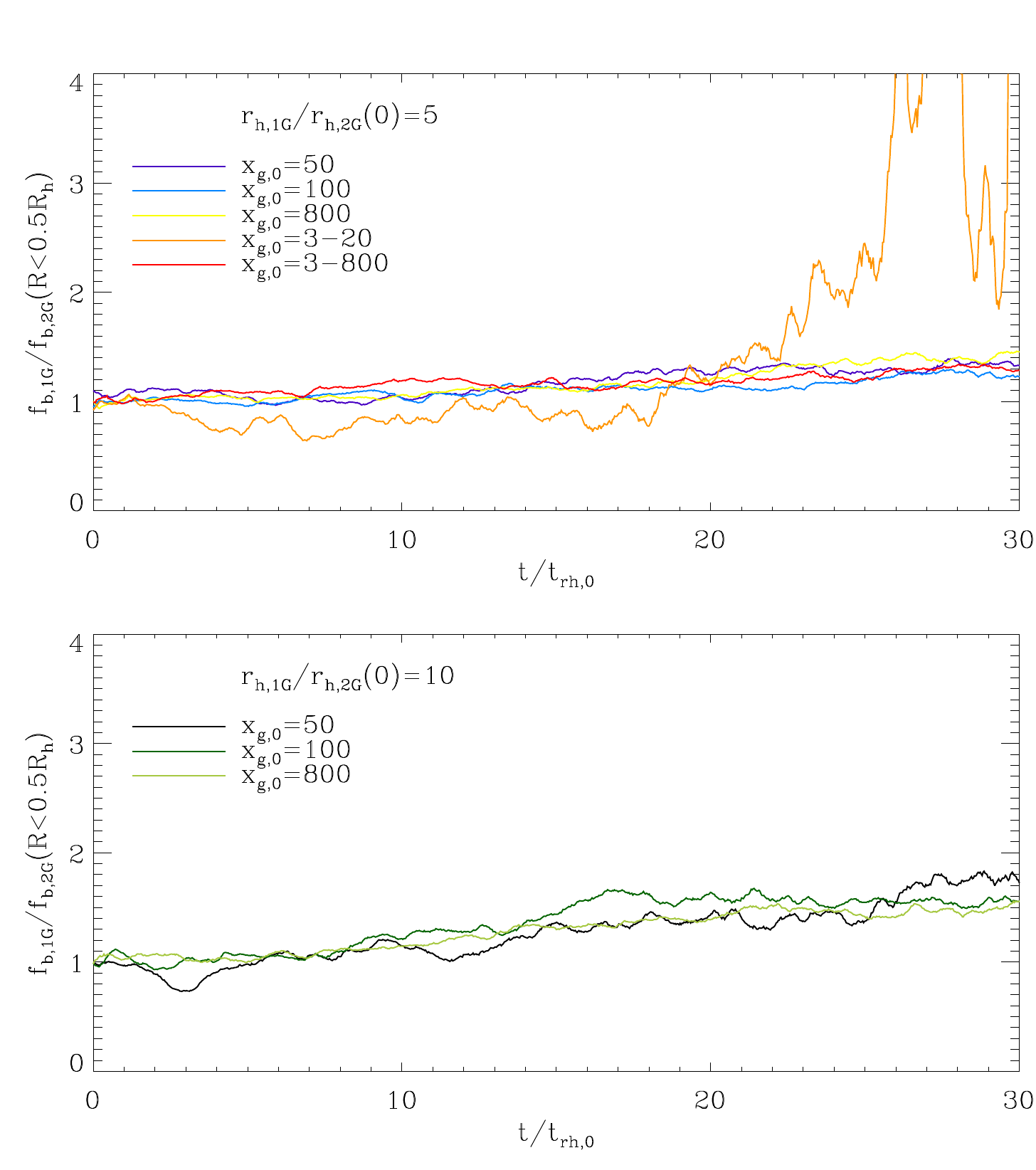}
  \caption{
  Time evolution of the ratio between 1G to the 2G binary incidences binaries calculated at projected distances smaller than 0.5$R_{\rm h}$. See text for details on the N-body simulations.
  }
 \label{fig:fractionratio} 
\end{figure} 
\end{centering}

In the study presented here, however, the {\it HST} data are limited to the inner regions between the clusters' centers and an outer radius ranging from about 0.3$R_{\rm h}$ to about 0.8 $R_{\rm h}$.
To further illustrate the expected dynamical effects on the evolution of the 1G and 2G binary incidence in the cluster's inner regions  we show in Figure~\ref{fig:fractionratio} the time evolution of the ratio $f^{\rm 1G-1G}_{\rm bin}/f^{\rm 2G-2G}_{\rm bin}$  measured between the cluster's center and 0.5 $R_{\rm h}$ for some of the simulations discussed  in \citet{hong2015a, hong2016a}. 
  Each simulation corresponds to different  
 values of X$_{\rm g,0}$, which is the parameter indicative of the initial hardness of primordial binaries. Specifically, X$_{\rm g,0} = E_{\rm b}/(m \sigma^{2})$, where $E_{\rm b}$ is the absolute values of the binary binding energy, and $\sigma$ is the 1D velocity dispersion of all stars. Upper and lower panels correspond to different ratios between the half light radii of 1G and 2G stars at formation.
 
 These figures clearly illustrate that the similar values of the 1G and 2G binary incidences found in our analysis are, in general, consistent with those expected in the cluster's innermost regions. 
  and in the outer regions of all the systems studied \citep[see e.g. Figure~12 in][]{hong2016a}.

Larger differences between the 1G and the 2G binary incidences are expected at all radial distances (including the inner regions) in systems with softer binaries and in the outer regions of all the systems studied \citep[see e.g. Figure~12 in][]{hong2016a}. The predicted increase in $f_{\rm b,1G}/f_{\rm b,2G}$ with the distance 
 from a cluster's center is consistent with what is found in previous studies based on radial velocities which probed the clusters' outer regions.
Specifically, \citet{lucatello2015a} analyzed multi-epoch spectra of 968 RGB stars of ten GCs and identified 21 radial-velocity binaries, corresponding
to a binary fraction of 2.2$\pm$0.5\%.  When they divided the stars into 
1G and 2G on the basis of their abundances of sodium and oxygen, they 
found that the fraction of binaries among 1G stars was 4.9$\pm$1.3\% and 
is significantly higher than the fraction of 2G binaries (1.2$\pm$0.4\%).

In another recent paper based on 384 stars of the GC NGC\,6362, \citet{dalessandro2018a} identified 12 binaries on the basis of their 
radial distribution, corresponding to a binary fraction of 3.1$\pm$0.9\%. 
When separating the stars into 1G and 2G on the basis of their sodium 
abundance, they find that only {\it one\,} binary belongs to the 
2G, implying a binary fraction of 0.7$\pm$0.7\%.  In contrast, the 
fraction of 1G binaries is significantly higher and corresponds 
to 4.7$\pm$1.4\%. Although a systematic study of the radial variation of the 1G and 2G binary incidences is necessary, the comparison between the similar values of $f_{\rm bin}^{\rm 1G-1G}$ and $f_{\rm bin}^{\rm 2G-2G}$ we find in the inner regions of this cluster and the larger 1G binary incidence found by \citet{dalessandro2018a} provides the first evidence of radial variation in the ratio of the 1G to the 2G binary incidences.

\begin{centering} 
\begin{figure} 
  \includegraphics[width=8.5cm,trim={0.0cm 0.0cm 6.2cm 13.0cm},clip]{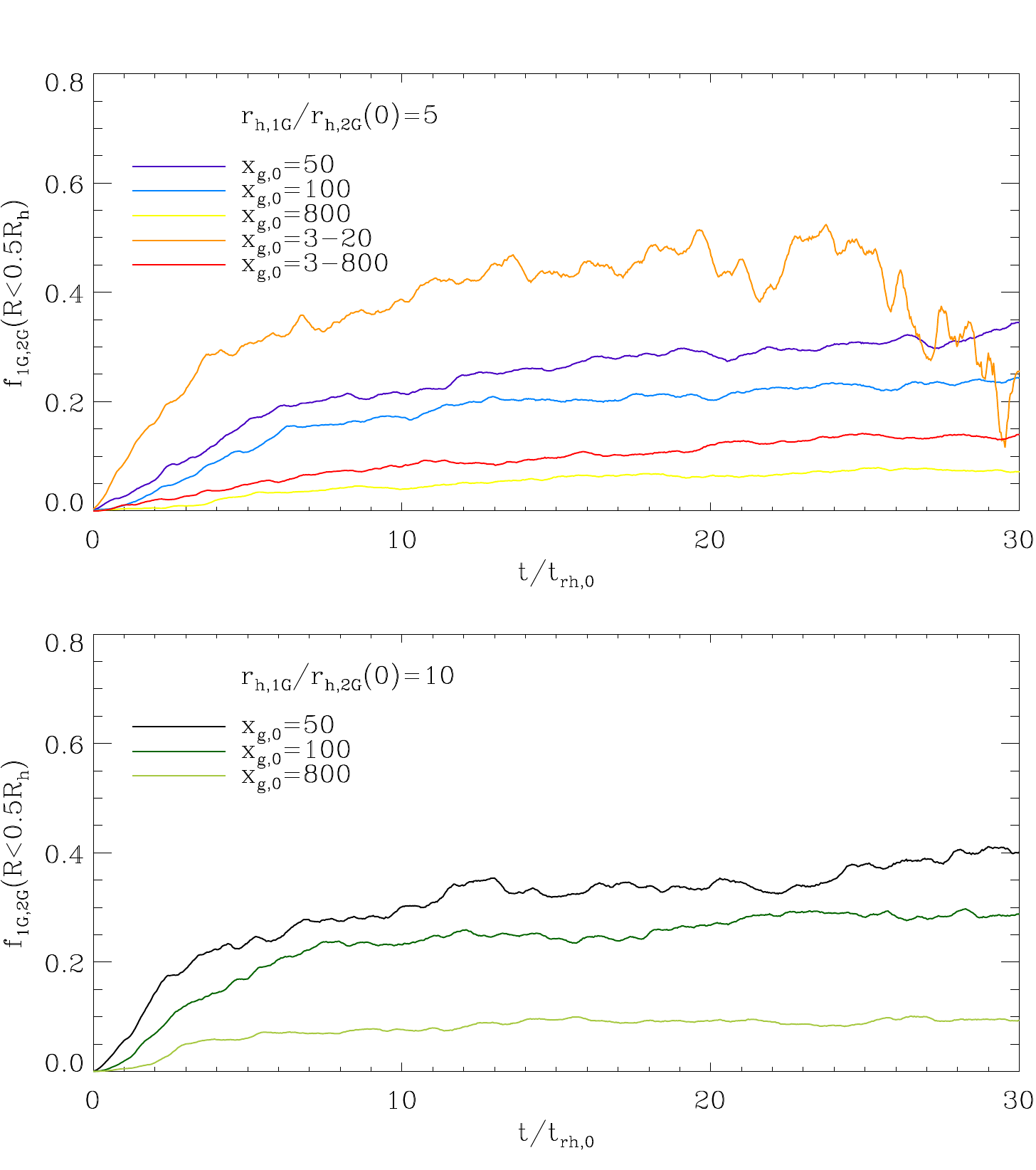}
  \caption{
  Time evolution of the fraction of 1G-2G binaries  (see text for details on the N-body simulations).}
 \label{fig:fractionmix} 
\end{figure} 
\end{centering}

In addition to the evolution of the fractions of 1G and 2G binaries, the simulations presented in \citet{hong2015a, hong2016a} predicted that exchange encounters during which one of the binary components can be replaced by one of the interacting stars can produce mixed binaries composed of one 1G star and one 2G star. The fraction of these binaries also depends on the cluster's dynamical age and the binary binding energy and provides a new and interesting tool to explore the dynamics of binary stars in multiple-population clusters \citep[see][for further discussion]{hong2015a, hong2016a}.
The photometric study presented in this paper has allowed us to reveal for the first time the presence of mixed binaries in NGC\,6352 at a statistical significance larger than 3$\sigma$, and suggest their presence in NGC\,288 and NGC\,6838 (at a confidence level of $\sim 2 \sigma$. Although more extensive observational and theoretical studies are needed, mixed binaries can provide an important insight in the binary dynamical activity in a cluster's inner regions.
Figure\,\ref{fig:fractionmix} shows the time evolution of the fraction of mixed binaries in the clusters' inner regions ($R<0.5R_{\rm h}$) for some of the simulations discussed in \citet{hong2015a, hong2016a} and illustrates the increase in the fraction of mixed binaries and its dependence on the binary binding energy for a few cases. In all cases the fraction of mixed binaries increases with time and is expected to be larger for denser clusters in a more advanced stage of their dynamical evolution and is expected to depend on the binary binding energy \citep[see also Figure~6 in][]{hong2016a}.

We emphasize again that these simulations are still idealized and not meant for a detailed comparison with observational  data; rather, the results  shed light on the dynamics driving the evolution of the 1G and 2G binary populations, the formation of mixed binaries, and illustrate the fundamental dynamical aspects behind the results emerging from our observational study. Additional numerical and observational studies will be needed to explore possible correlations between between 1G, 2G and mixed binary properties, the present-day cluster structural properties and the cluster's dynamical history. 

\section{Summary and conclusions}\label{sec:conclusions}
In this analysis, we used {\it HST} data collected within the UV survey of Galactic GCs \citep{piotto2015a} to investigate the incidence of binaries in five GCs by using multi-band photometry. We used the $m_{\rm F336W}-m_{\rm F438W}$ vs.\,$m_{\rm F275W} - m_{\rm F336W}$ two-color diagrams and the ChM to identify 1G and 2G stars along the RGB, SGB and MS of each cluster. 
We selected a sample of binaries from the optical $m_{\rm F814W}$ vs.\,$m_{\rm F606W} − m_{\rm F814W}$ CMD, which are composed of pairs of stars with similar luminosity and derived their distribution in the $m_{\rm F814W}$ vs.\,$C_{\rm F275W,F336W,F438W}$ pseudo CMD. We compared the $C_{\rm F275W,F336W,F438W}$ pseudo-color distribution of the observed binaries with the corresponding distribution of a large sample of simulated stellar populations that include various combinations of 1G- 1G, 1G-2G and 2G-2G stars.

We find that in NGC\,288, NGC\,6352, NGC\,6362 and NGC\,6838 the incidence of 1G-1G binaries among 1G star is similar to the incidence of binaries among 2G-2G stars.
M\,4, where the fraction of 1G-1G binary pairs among 1G stars is 3.1$\pm$0.9 times higher than the fraction of 2G-2G binaries among 2G stars, is a remarkable exception.

The method presented in this paper, makes it possible to identify for the first time mixed 1G-2G binary systems, binaries composed of one 1G star and one 2G star.
N-body simulations predicted mixed binaries to form in binary interactions during which one binary component is replaced by one of the interacting stars of a different population. These binaries provide a new tool to explore binary activity and dynamical history of multiple stellar populations.

 While a statistically-significant detection has been found only in NGC\,6352, at face value the best fit fraction of 1G-2G binaries is smaller than $\sim$0.15 in NGC\,288, NGC\,6121 and NGC\,6362, whereas NGC\,6838 and NGC\,6352 host larger fractions of 1G-2G binaries ($\sim$0.27 and $\sim$0.48).
Using the fraction of mixed binaries we provided an initial estimate of the fraction of the observed binary population consistent with being primordial and not the results of exchange interactions and/or dynamical binary formation. Although additional investigation of this issue is needed, our initial estimates suggest that most binaries in NGC\,6121 and NGC\,6362 are consistent with a primordial origin, while in NGC\,6352 most binaries could be the result of dynamical interactions. In NGC\,6838 and NGC\,288 the number of binaries with a primordial origin is similar to that of dynamically formed binaries.
Future studies extending the analysis presented here to a larger sample of clusters and probing a broader range of radial distances from a cluster's  center will be necessary to build a complete picture of the dynamical effects on binaries in multiple-population globular clusters and provide new constraints for theoretical studies of the formation and evolution of multiple populations.

\section*{acknowledgments} 
\small
We thank Antonio Sollima for several suggestions that improved the manuscript. 
This work has received funding from the European Research Council (ERC) 
under the European Union's Horizon 2020 research innovation programme 
(Grant Agreement ERC-StG 2016, No 716082 `GALFOR', PI: Milone, http://progetti.dfa.unipd.it/GALFOR),  and the European Union's Horizon 2020 research and innovation programme 
under the Marie Sklodowska-Curie (Grant Agreement No 797100, beneficiary: 
Marino). APM  and MT acknowledge support from MIUR through the FARE project 
R164RM93XW SEMPLICE (PI: Milone). APM and LRB acknowledge support by MIUR
under PRIN program 2017Z2HSMF (PI: Bedin).
DN acknowledges partial support by the Universit\'a degli Studi di Padova Progetto di Ateneo BIRD178590.

\bibliography{ms} 
\end{document}